\documentclass[useAMS,usenatbib,twocolumn]{mn2e}
\usepackage{times} 
\usepackage{amsmath}
\usepackage{rotating}
\usepackage{amssymb}

\def\apj{ApJ}

\def\apjl{ApJ}
\def\mnras{MNRAS}

\def\physrep{Phys. Rep.}
\def\nat{Nature}

\def\apjs{ApJS}
\def\prd{Phys. Rev. D}

\title[Phase-space structure in the local dark matter distribution]
      {Phase-space structure in the local dark matter distribution and
        its signature in direct detection experiments}
      \author[Vogelsberger et al.] {\parbox{17.5cm}{ Mark
          Vogelsberger$^{1}$, Amina Helmi$^{2}$, Volker
          Springel$^{1}$, Simon~D.~M.~White$^{1}$, \\ Jie Wang$^{1,3}$, 
          Carlos~S.~Frenk$^{3}$, Adrian Jenkins$^{3}$, Aaron Ludlow$^{2}$,
          Julio ~F.~Navarro$^{2,5}$        
        }\vspace{0.3cm}\\ $^1$Max-Planck-Institut f\"{u}r Astrophysik,
        Karl-Schwarzschild-Stra\ss{}e 1, 85740 Garching bei
        M\"{u}nchen, Germany\\ $^{2}${Kapteyn Astronomical Institute,
          Univ. of Groningen, P.O. Box 800, 9700 AV Groningen, The
          Netherlands}\\ $^{3}${Institute for Computational Cosmology,
          Dep. of Physics, Univ. of Durham, South Road, Durham DH1
          3LE, UK}\\ $^{4}${Dep. of Physics \& Astron., University of
          Victoria, Victoria, BC, V8P 5C2, Canada}\\ $^{5}${Department
          of Astronomy, University of Massachusetts, Amherst, MA
          01003-9305, USA}\\ }

\begin{document}
\date{Accepted ???. Received ???; in original form ???}

\pagerange{\pageref{firstpage}--\pageref{lastpage}} \pubyear{2008}

\maketitle

\label{firstpage}

\begin{abstract}
We study predictions for dark matter phase-space structure near the
Sun based on high-resolution simulations of six galaxy halos taken
from the Aquarius Project. The local DM density distribution is
predicted to be remarkably smooth; the density at the Sun differs from
the mean over a best-fit ellipsoidal equidensity contour by less than
$15\%$ at the $99.9\%$ confidence level. The local velocity
distribution is also very smooth, but it differs systematically from a
(multivariate) Gaussian distribution. This is not due to the presence
of individual clumps or streams, but to broad features in the velocity
modulus and energy distributions that are stable both in space and
time and reflect the detailed assembly history of each halo. These
features have a significant impact on the signals predicted for WIMP
and axion searches. For example, WIMP recoil rates can deviate by
$\sim 10\%$ from those expected from the best-fit multivariate
Gaussian models. The axion spectra in our simulations typically peak
at lower frequencies than in the case of multivariate Gaussian
velocity distributions. Also in this case, the spectra show
significant imprints of the formation of the halo. This implies that
once direct DM detection has become routine, features in the detector
signal will allow us to study the dark matter assembly history of the
Milky Way.  A new field, ``dark matter astronomy'', will then emerge.

\end{abstract}


\section{Introduction}

In the 75 years since \cite{1933AcHPh...6..110Z} first pointed out the need for
substantial amounts of unseen material in the Coma cluster, the case
for a gravitationally dominant component of non-baryonic dark matter
has become overwhelmingly strong. It seemed a long shot when
\cite{1982ApJ...263L...1P} first suggested that the dark matter might be an
entirely new, weakly interacting, neutral particle with very low
thermal velocities in the early universe, but such Cold Dark Matter (CDM)
is now generally regarded as the most plausible and consistent
identification for the dark matter.  Particle physics has suggested
many possible CDM particles beyond the standard model. Two promising
candidates are WIMPs \citep[weakly interacting massive particles,
  see][]{1977PhRvL..39..165L, 1978ApJ...223.1015G,1984STIN...8526469E}
and axions \citep{1977PhRvL..38.1440P, 1977PhRvD..16.1791P,
  1978PhRvL..40..223W, 1978PhRvL..40..279W}. Among the WIMPs, the
lightest supersymmetric particle, the neutralino, is currently favoured
as the most likely CDM particle, and the case will be enormously
strengthened if the LHC confirms supersymmetry. However, ultimate
confirmation of the CDM paradigm can only come through the direct or
indirect detection of the CDM particles themselves.  Neutralinos, for
example, are their own antiparticles and can annihilate to produce
$\gamma$-rays and other particles. One goal of the recently launched 
Fermi Gamma-ray space telescope is to detect this radiation
\citep{1999APh....11..277G,2008Natur.456...73S}.  

Direct detection experiments, on the other hand, search for the interaction of CDM
particles with laboratory apparatus.  For WIMPs, detection is based on
nuclear recoil events in massive, cryogenically cooled bolometers in
underground laboratories \citep{1996PhR...267..195J}; for axions,
resonant microwave cavities in strong magnetic fields exploit the
axion-photon conversion process \citep{1985PhRvD..32.2988S}.  Despite
intensive searches, the only experiment which has so far reported a
signal is DAMA \citep[][]{2007hepi.conf..214B} which has clear evidence for an annual
modulation of their event rate of the kind expected from the Earth's
motion around the Sun. The interpretation of this result is
controversial, since it appears to require dark matter properties
which are in conflict with upper limits established by other
experiments
\citep[see][for a discussion and possible solutions]{2004PhRvD..70l3513S,2005PhRvD..71l3520G,2006JPhCS..39..166G}. 
Regardless of this, recent improvements in
detector technology may enable a detection of ``standard model'' WIMPS
or axions within a few years.

Event rates in all direct detection experiments are determined by the
local DM phase-space distribution at the Earth's position. The
relevant scales are those of the apparatus and so are extremely small
from an astronomical point of view. As a result, interpreting null
results as excluding specific regions of candidate parameter space
must rely on (strong) assumptions about the fine-scale structure of
phase-space in the inner Galaxy. In most analyses the dark matter has
been assumed to be smoothly and spherically distributed about the
Galactic Centre with an isotropic Maxwellian velocity distribution
\citep[e.g.][]{1988PhRvD..37.3388F} or a multivariate Gaussian
distribution \citep[e.g.][]{2001JHEP...03..049U, 2001PhRvD..63d3005G,
  2002PhRvD..66f3502H}.  The theoretical justification for these
assumptions is weak, and when numerical simulations of halo formation
reached sufficiently high resolution, it became clear that the
phase-space of CDM halos contains considerable substructure, both
gravitationally bound subhalos and unbound streams. As numerical
resolution has improved it has become possible to see structure closer
and closer to the centre, and this has led some investigators to
suggest that the CDM distribution near the Sun could, in fact, be
almost fractal, with large density variations over short length-scales
\citep[e.g.][]{2008PhRvD..77j3509K}. This would have substantial consequences
for the ability of direct detection experiments to constrain particle
properties.

Until very recently, simulation studies were unable to resolve any
substructure in regions as close to the Galactic Centre as the Sun
\citep[see][for example]{2001PhRvD..64f3508M,2002PhRvD..66f3502H,
  2003MNRAS.339..834H}. This prevented realistic evaluation of the
likelihood that massive streams, clumps or holes in the dark matter
distribution could affect event rates in Earth-bound detectors and so
weaken the particle physics conclusions that can be drawn from null
detections \citep[see][for recent
  discussions]{2006PhRvD..74d3531S,2008PhRvD..77j3509K}.  As we shall
show in this paper, a new age has dawned. As part of its Aquarius
Project \citep{2008arXiv0809.0898S} the Virgo Consortium has carried out a
suite of ultra-high resolution simulations of a series of Milky
Way-sized CDM halos. Simulations of individual Milky Way halos of
similar scale have been carried out by \cite{2008Natur.454..735D} and \cite{2008arXiv0808.2981S}. 
Here we use the Aquarius simulations to provide the first
reliable characterisations of the local dark matter phase-space
distribution and of the detector signals which should be anticipated
in WIMP and axion searches.  

\begin{figure}
\center{
\includegraphics[width=0.4\textwidth]{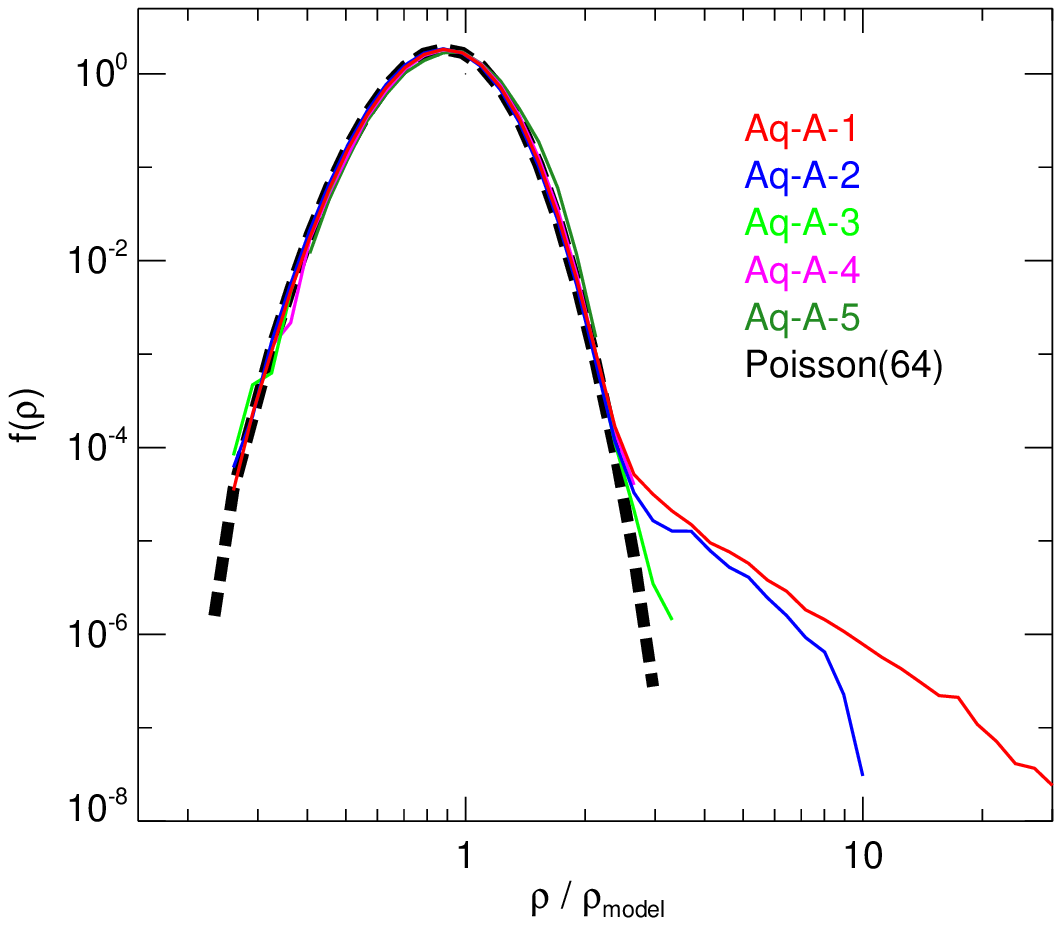}
\includegraphics[width=0.4\textwidth]{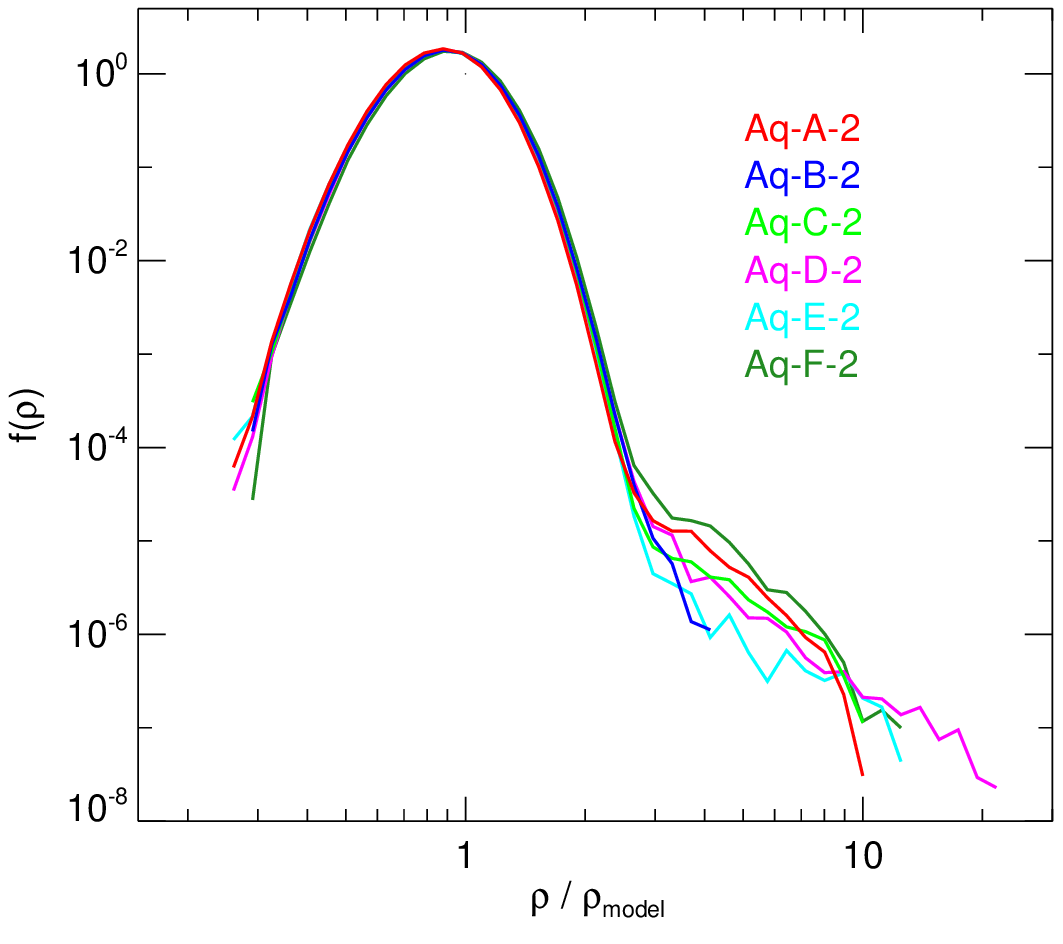}
}
\caption{Top panel: Density probability distribution function (DPDF)
  for all resimulations of halo Aq-A measured within a thick
  ellipsoidal shell between equidensity surfaces with major axes of $6$
  and $12$~kpc. The local dark matter density at the position of each
  particle, estimated using an SPH smoothing technique, is divided by
  the density of the best-fit, ellipsoidally stratified, power-law
  model. The DPDF gives the distribution of the local density in units
  of that predicted by the smooth model at random points within the
  ellipsoidal shell. At these radii only resolution levels 1 and 2 are
  sufficient to follow substructure.  As a result, the characteristic
  power-law tail due to subhalos is not visible at lower
  resolution. The fluctuation distribution of the smooth component is
  dominated by noise in our 64-particle SPH density estimates. The
  density distribution measured for a {\it uniform} (Poisson) particle
  distribution is indicated by the black dashed line.  Bottom panel:
  As above, but for all level-2 halos after rescaling to $V_{\rm max}=208.49$~km/s. 
  In all cases the core of the DPDF is dominated
  by measurement noise and the fraction of points in the power law
  tail due to subhalos is very small. The chance that the Sun lies
  within a subhalo is $\sim 10^{-4}$. With high probability the local
  density is close to the mean value averaged over the Sun's
  ellipsoidal shell. }
\label{fig:DPDF} 
\end{figure}

\section{The Numerical Simulations}
 
The cosmological parameters for the Aquarius simulation set are
$\Omega_m=0.25,\Omega_\Lambda=0.75,\sigma_8=0.9,n_s=1$ and
$H_0=100~h~{\rm km}~{\rm s}^{\rm -1}~{\rm Mpc}^{\rm -1}$ with
$h=0.73$, where all quantities have their standard definitions.  These
parameters are consistent with current cosmological constraints within
their uncertainties, in particular, with the parameters inferred from
the WMAP 1-year and 5-year data analyses \citep{2003ApJS..148..175S, 
2008arXiv0803.0547K}. Milky Way-like halos were selected for resimulation from a parent
cosmological simulation which used $900^3$ particles to follow the
dark matter distribution in a $100 h^{-1} {\rm Mpc}$ periodic box.
Selection was based primarily on halo mass ($\sim 10^{12} M_\odot$)
but also required that there should be no close and massive neighbour
at $z=0$. The Aquarius Project resimulated six such halos at a series
of higher resolutions.  The naming convention uses the tags Aq-A
through Aq-F to refer to these six halos. An additional suffix $1$ to
$5$ denotes the resolution level. Aq-A-1 is the highest resolution
calculation, with a particle mass of $1.712 \times 10^3~M_\odot$ and a
virial mass of $1.839 \times 10^{12}~M_\odot$ it has more than a
billion particles within the virial radius $R_{\rm 200}$ which we
define as the radius containing a mean density 200 times the critical
value.  The Plummer equivalent softening length of this run is
$20.5$~pc. Level-2 simulations are available for all six halos with
about $200$ million particles within $R_{\rm 200}$.  Further details
of the halos and their characteristics can be found in
\cite{2008arXiv0809.0898S}.

In the following analysis we will often compare the six level-2
resolution halos, Aq-A-2 to Aq-F-2. To facilitate this comparison, we
scale the halos in mass and radius by the constant required to give
each a maximum circular velocity of $V_{\rm max}=208.49$~km/s, the
value for Aq-A-2. We will also sometimes refer to a coordinate system
that is aligned with the principal axes of the inner halo, and which
labels particles by an ellipsoidal radius $r_{\rm ell}$ defined as the semi-major
axis length of the ellipsoidal equidensity surface on which the
particle sits. We determine the orientation and shape of these
ellipsoids as follows. For each halo we begin by diagonalising the
moment of inertia tensor of the dark matter within the spherical shell
$6~{\rm kpc}<r<12~{\rm kpc}$ (after scaling to a common $V_{\rm
  max}$). This gives us a first estimate of the orientation and shape
of the best fitting ellipsoid.  We then reselect particles with
$6~{\rm kpc}<r_{\rm ell}<12~{\rm kpc}$, recalculate the moment of inertia
tensor and repeat until convergence. The resulting ellipsoids have
minor-to-major axis ratios which vary from $0.39$ for Aq-B-2 to $0.59$ for
Aq-D-2. The radius restriction reflects our desire to probe the dark
matter distribution near the Sun.

\section{Spatial distributions}

The density of DM particles at the Earth determines the flux of DM
particles passing through laboratory detectors. It is important,
therefore, to determine not only the mean value of the DM density 8~kpc
from the Galactic Centre, but also the fluctuations around this mean
which may result from small-scale structure.

\begin{figure}
\center{
\includegraphics[width=0.5\textwidth]{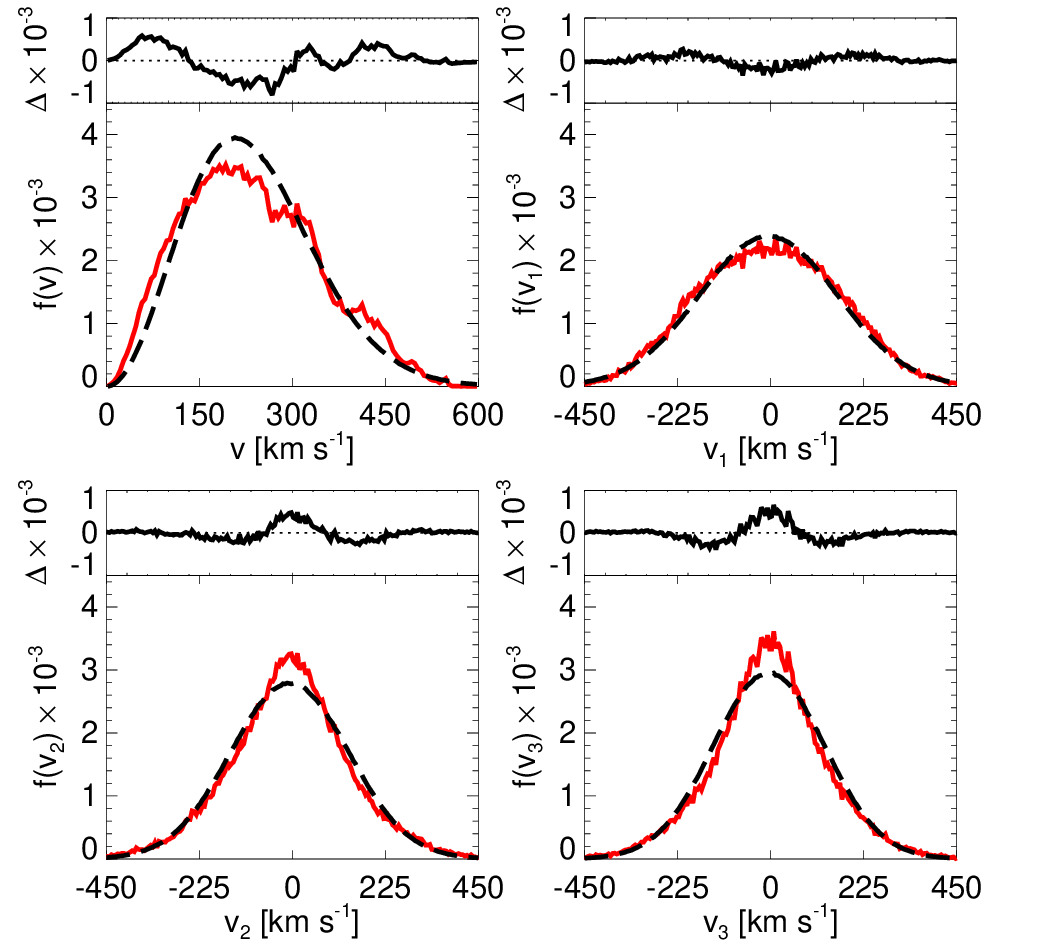}
\includegraphics[width=0.45\textwidth]{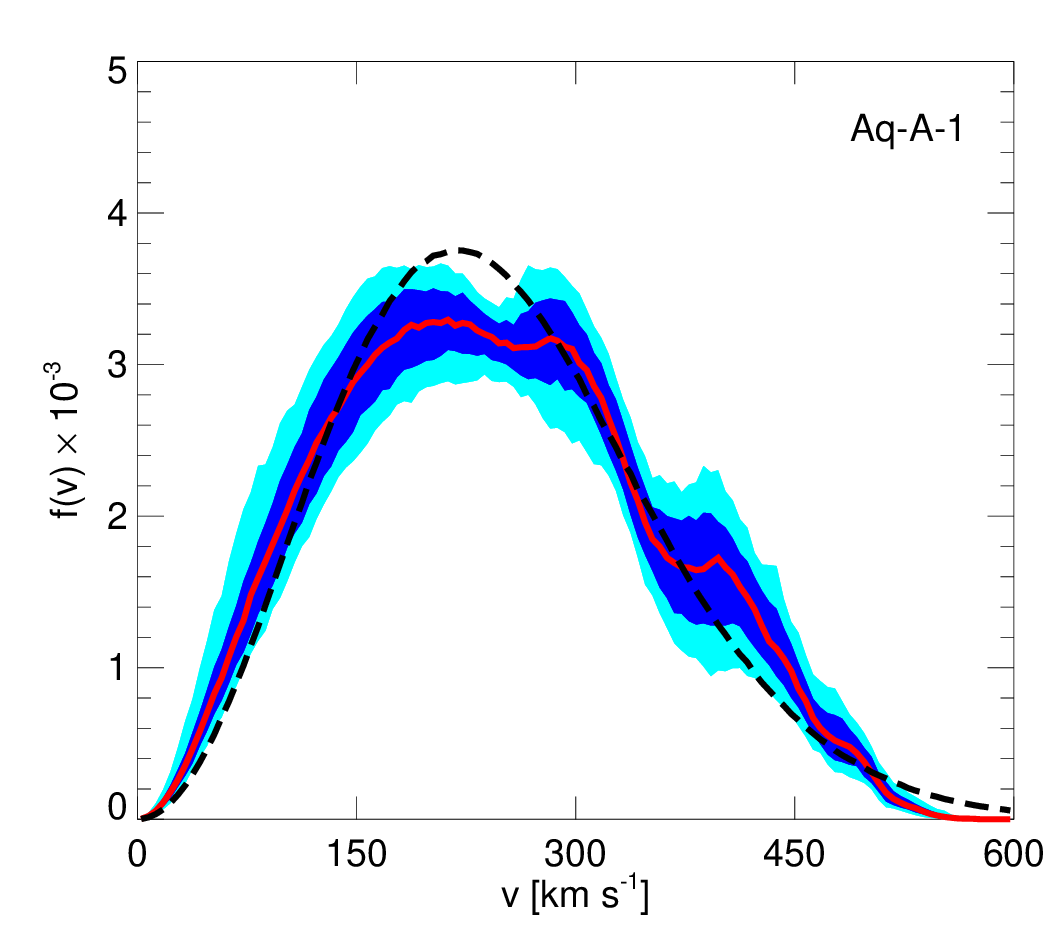}
}
\caption{Top four panels: Velocity distributions in a $2$~kpc box at
  the Solar Circle for halo Aq-A-1. $v_1$, $v_2$ and $v_3$ are the
  velocity components parallel to the major, intermediate and minor
  axes of the velocity ellipsoid; $v$ is the modulus of the velocity
  vector.  Red lines show the histograms measured directly from the
  simulation, while black dashed lines show a multivariate Gaussian
  model fit to the individual component distributions. Residuals from
  this model are shown in the upper part of each panel. The major axis
  velocity distribution is clearly platykurtic, whereas the other two
  distributions are leptokurtic. All three are very smooth, showing no
  evidence for spikes due to individual streams. In contrast, the
  distribution of the velocity modulus, shown in the upper left panel,
  shows broad bumps and dips with amplitudes of up to ten percent of
  the distribution maximum. Lower panel: Velocity modulus
  distributions for all $2$~kpc boxes centred between $7$ and $9$~kpc
  from the centre of Aq-A-1. At each velocity a thick red line gives
  the median of all the measured distributions, while a dashed black
  line gives the median of all the fitted multivariate Gaussians.  The
  dark and light blue contours enclose $68 \%$ and $95 \%$ of all the
  measured distributions at each velocity. The bumps seen in the
  distribution for a single box are clearly present with similar
  amplitude in all boxes, and so also in the median curve. The bin
  size is $5$~km/s in all plots.}
\label{fig:BoxVelA} 
\end{figure}

We estimate the local DM distribution at each point in our simulations
using an SPH smoothing kernel adapted to the 64 nearest neighbours.
We then fit a power law to the resulting distribution of $\ln\rho$
against $\ln r_{\rm ell}$ over the ellipsoidal radius range $6~{\rm
  kpc}<r_{\rm ell}<12~{\rm kpc}$. This defines a smooth model density
field $\rho_{\rm model}(r_{\rm ell})$. We then construct a density probability
distribution function (DPDF) as the histogram of $\rho/\rho_{\rm model}$
for all particles in $6~{\rm kpc}<r_{\rm ell}<12~{\rm kpc}$, where each is
weighted by $\rho^{-1}$ so that the resulting distribution refers to
random points within our ellipsoidal shell rather than to random mass
elements. We normalise the resulting DPDFs to have unit integral.
They then provide a probability distribution for the local dark matter
density at a random point in units of that predicted by the best
fitting smooth ellipsoidal model.

In Fig.~\ref{fig:DPDF} we show the DPDFs measured in this way for all
resimulations of Aq-A (top panel) and for all level-2 halos after
scaling to a common $V_{\rm max}$ (bottom panel).  Two distinct
components are evident in both plots. One is smoothly and log-normally
distributed around $\rho=\rho_{\rm model}$, the other is a power-law
tail to high densities which contains less than $10^{-4}$ of all
points.  The power-law tail is not present in the lower resolution
halos (Aq-A-3, Aq-A-4, Aq-A-5) because they are unable to resolve
subhalos in these inner regions. However, Aq-A-2 and Aq-A-1 give quite
similar results, suggesting that resolution level 2 is sufficient to
get a reasonable estimate of the overall level of the tail. A
comparison of the six level 2 simulations then demonstrates that this
tail has similar shape in different halos, but a normalisation which
can vary by a factor of several. In none of our halos does the fraction
of the distribution in this tail rise above $5\times
10^{-5}$. Furthermore, the arguments of Springel et al (2008) suggest
that the total mass fraction in the inner halo (and thus also the
total volume fraction) in subhalos below the Aq-A-1 resolution limit
is at most about equal to that above this limit. Hence, the chance
that the Sun resides in a bound subhalo of any mass is of order
$10^{-4}$.

The striking similarity of the smooth log-normal component in all the
distributions of Fig.~\ref{fig:DPDF} has nothing to do with actual
density variations in the smooth dark matter distribution. It is, in
fact, simply a reflection of the noise in our local density estimates.
We demonstrate this by setting up a uniform Poisson point distribution
within a periodic box and then using an SPH smoothing kernel adapted
to the 64 nearest neighbours to associate a local density with each
particle in exactly the same way as for our halo simulations. We can
then construct a DPDF for these estimates (relative to their mean) in
exactly the same way as before. The result is shown in the top panel
of Fig.~\ref{fig:DPDF} as a dashed black line. It is an almost perfect
fit to the smooth component in the simulations, and it would fit the
other halos equally well if plotted in the lower panel.

\begin{figure}
\center{
\includegraphics[width=0.45\textwidth]{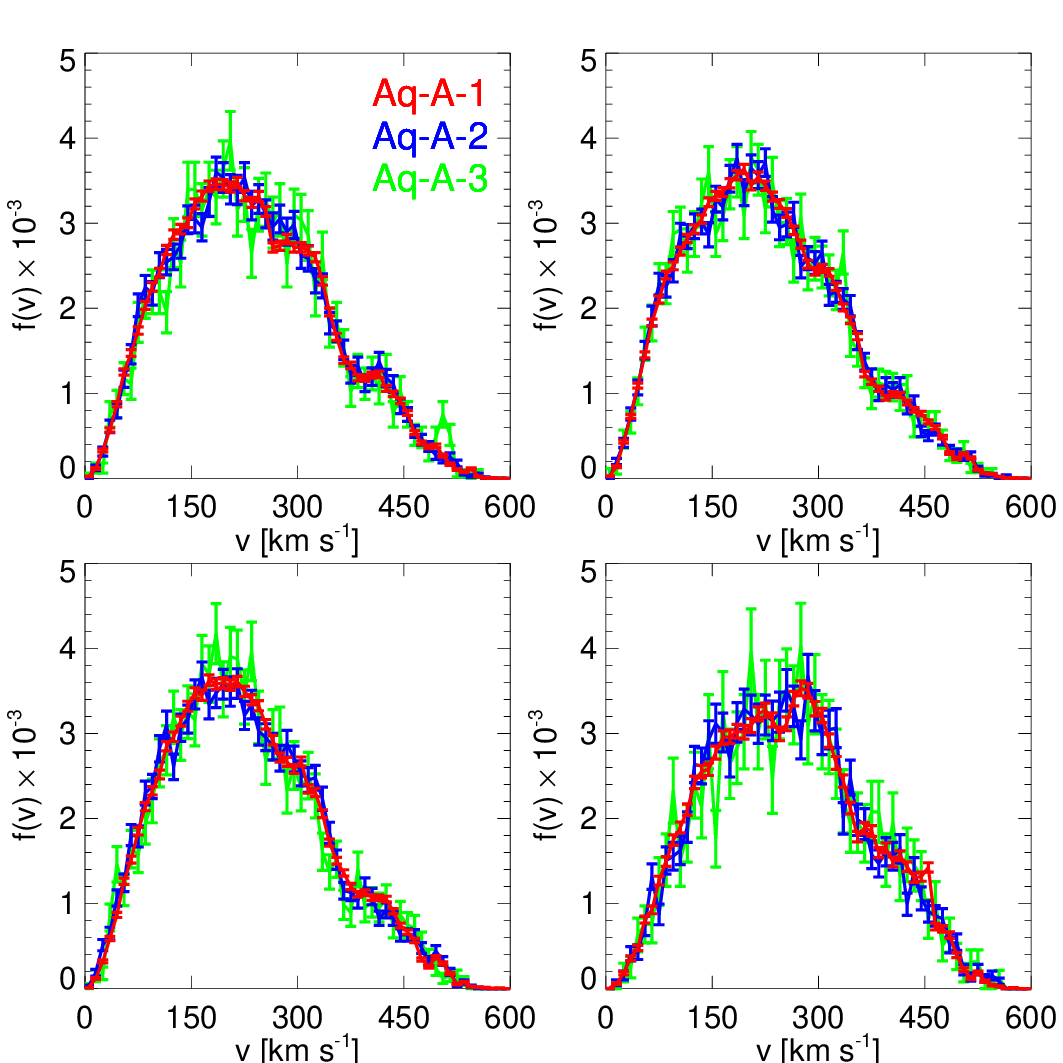}
}
\caption{Distributions of the velocity modulus in four well separated
  $2$~kpc boxes about 8~kpc from the centre of Aq-A. Results are shown
  for each region from each of the three highest resolution
  simulations.  Error bars are based on Poisson statistics. The
  different resolutions agree within their error bars, and show the
  same bumps in all four boxes. For the purpose of this plot, we have
  chosen a larger bin for our histograms, $10$~km/s as compared to
  5~km/s in our other velocity plots. For this bin size the
  statistical noise in Aq-A-1 is barely visible.
  \label{fig:GaussianBoxVel_Resolution} }
\end{figure}

The fit is not perfect, however, and it is possible to disentangle the
true scatter in density about the smooth model from the estimation
noise. The latter is expected to be asymptotically log-normal for
large neighbour numbers, and Fig.~\ref{fig:DPDF} shows that it is very
close to log-normal for our chosen parameters.  If we assume that the
scatter in intrinsic density about the smooth model is also
approximately log-normal, we can estimate its scatter as the square
root of the difference between the variance of the simulation scatter
and that of the noise: symbolically, $\sigma_{\rm intr} =
\sqrt{\sigma_{\rm obs}^2 - \sigma_{\rm noise}^2}$.  Indeed, it turns out that
the variance in $\ln (\rho/\rho_{\rm model})$ which we measure for our
simulated halos (excluding the power-law tail) is consistently higher
than that which we find for our uniform Poisson
distribution. Furthermore, tests show that the differences are stable
if we change the number of neighbours used in the SPH estimator to 32
or 128, even though this changes the noise variance by factors of
two. This procedure give the following estimates for {\it rms}
intrinsic scatter around the smooth model density field in our six
level-2 halos, Aq-A-2 to Aq-F-2: $2.2\%$, $4.4\%$, $3.7\%$, $2.1\%$,
$4.9\%$ and $4.0\%$ respectively. The very large particle number in
the radial range we analyse results in a standard error on these
number which is well below $0.01$ for all halos.  Thus, we can say
with better than $99.9\%$ confidence that the DM density at the Sun's
position differs by less than $15\%$ from the average over the
ellipsoidal shell on which the Sun sits. This small scatter implies
that the density field in the inner halo is remarkably well described
by a smooth, ellipsoidal, power-law model.

We conclude that the local density distribution of dark matter should
be very smooth. Bound clumps are very unlikely to have any effect on
direct detection experiments.  The main reason for this is the short
dynamical time at the solar radius (about 1\% of the Hubble
time). This results in very efficient mixing of unbound material and
the stripping of all initially bound objects to a small fraction of
the maximum mass they may have had in the past \citep[see][for a
  discussion of these processes]{2008MNRAS.385..236V}. Note that the
actual density of DM in the Solar neighbourhood and the shape of the
equidensity surfaces of the Milky Way's dark matter distribution will
depend on how the gravitational effects of the baryonic components
have modified structure during the system's formation. Unfortunately,
the shape of the inner DM halo of the Milky Way is poorly constrained
observationally \citep{2004ApJ...610L..97H,2005ApJ...619..807L}. The
dissipative contraction of the visible components probably increased
the density of the dark matter component, and may also have made it
more axisymmetric \citep[e.g.][]{2004ApJ...616...16G,2004ApJ...611L..73K} but 
these processes are unlikely to affect the level of small-scale structure. 
The very smooth behaviour we find in our pure dark matter halos should 
apply also to the more complex real Milky Way.

\section{Velocity distributions}

The velocity distribution of DM particles near the Sun is also an
important factor influencing the signal expected in direct detection
experiments. As mentioned in the Introduction, most previous work has
assumed this distribution to be smooth, and either Maxwellian or
multivariate Gaussian. Very different distributions are possible in
principle. For example, if the local density distribution is a
superposition of a relatively small number of DM streams, the local
velocity distribution would be effectively discrete with all particles
in a given stream sharing the same velocity
\citep{1995PhRvL..75.2911S, 2001PhRvD..64h3516S, 2003PhRvL..90u1301S}.
Clearly, it is important to understand whether such a distribution is
indeed expected, and whether a significant fraction of the local mass
density could be part of any individual stream.
\begin{figure*}
\center{
\includegraphics[width=0.4\textwidth]{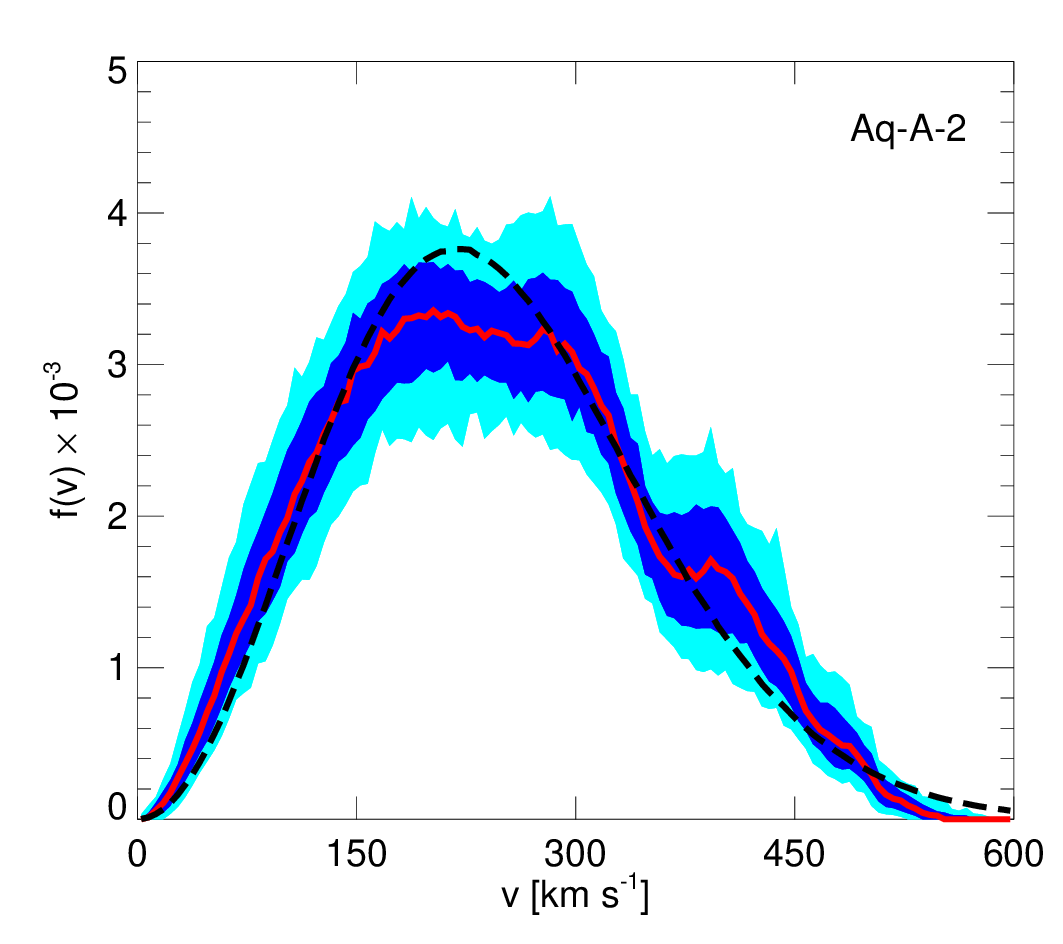}
\includegraphics[width=0.4\textwidth]{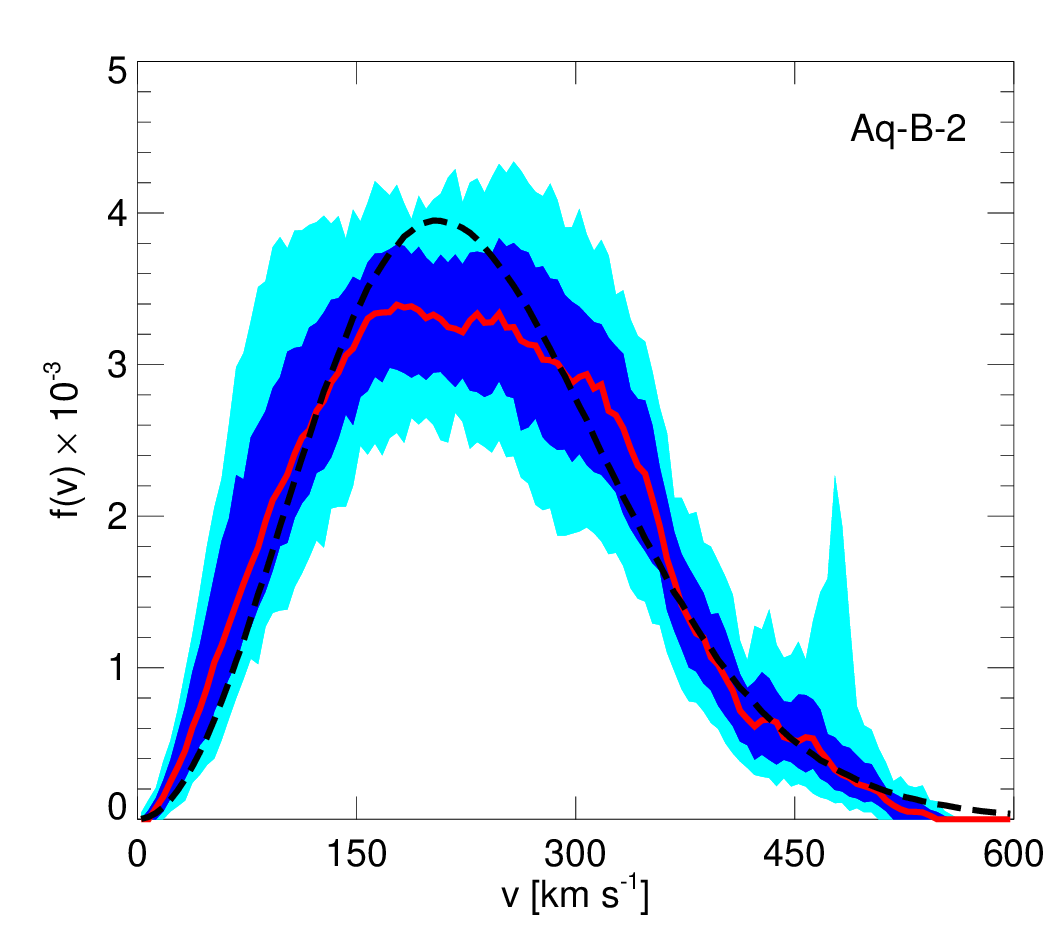}
\includegraphics[width=0.4\textwidth]{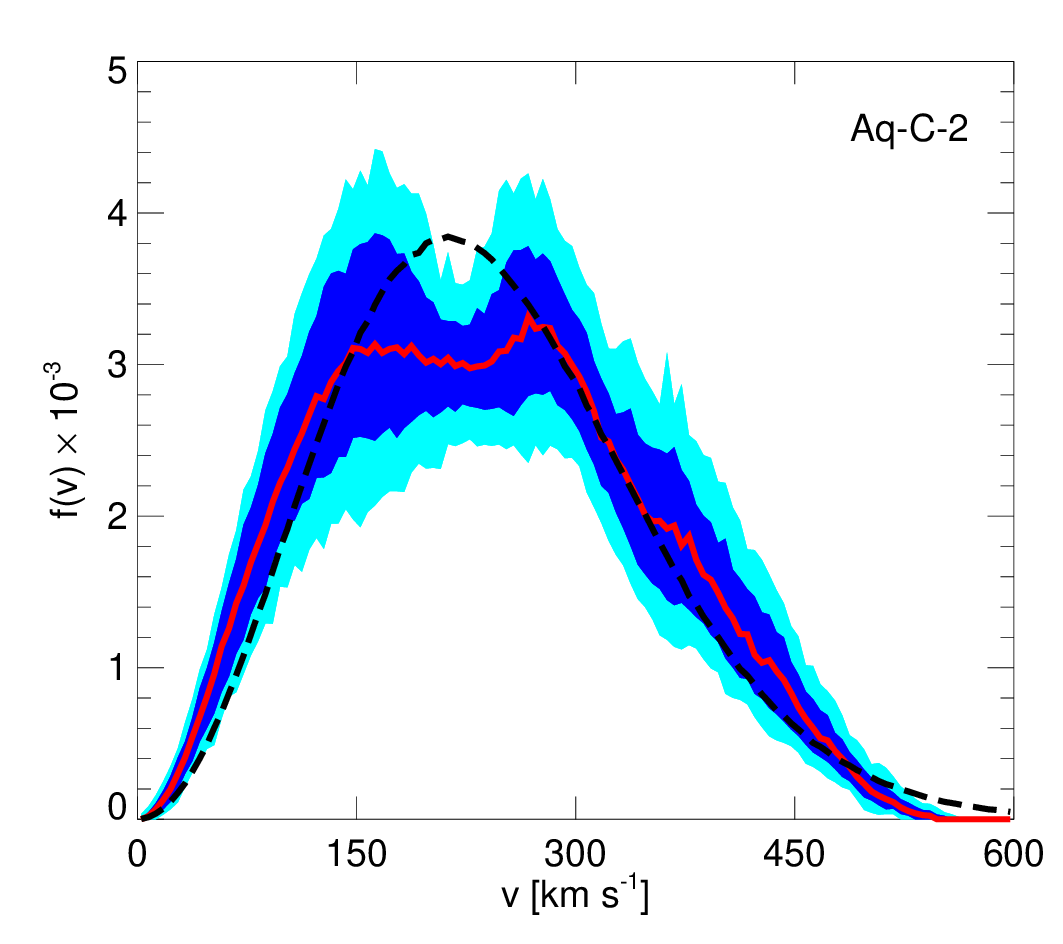}
\includegraphics[width=0.4\textwidth]{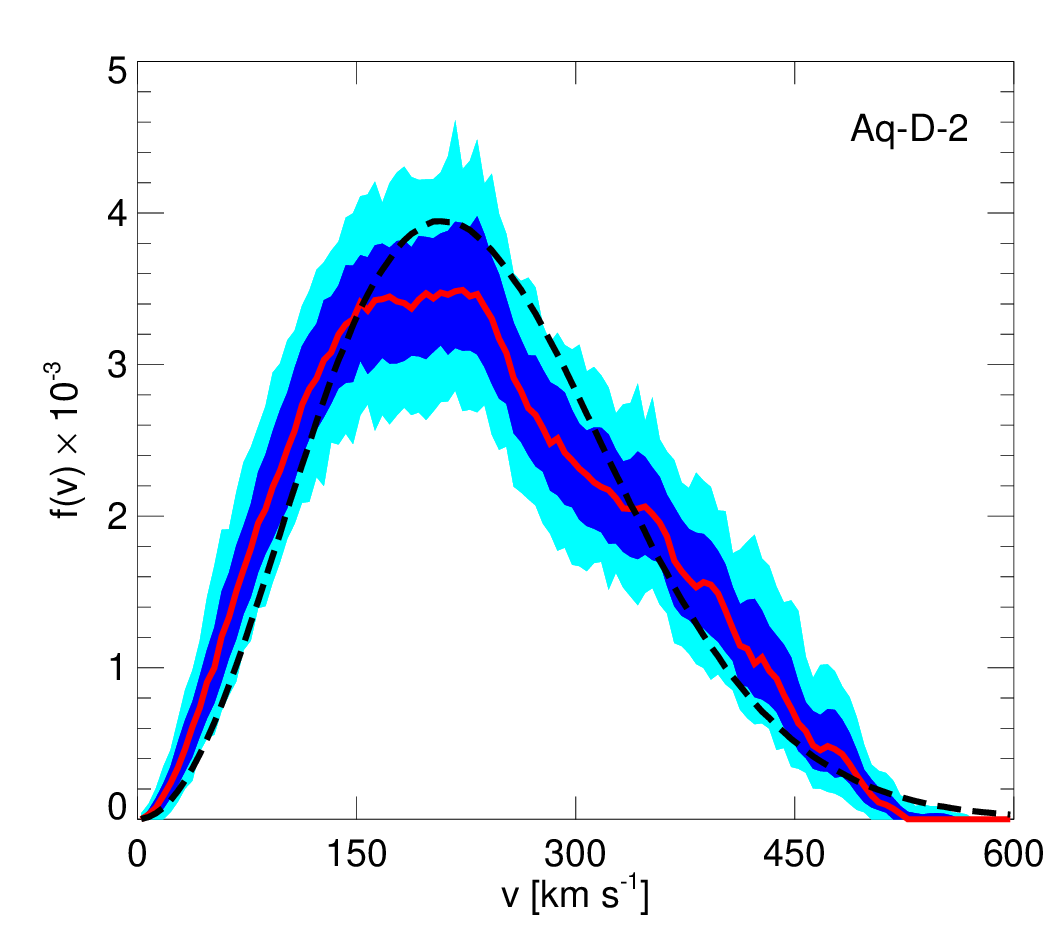}
\includegraphics[width=0.4\textwidth]{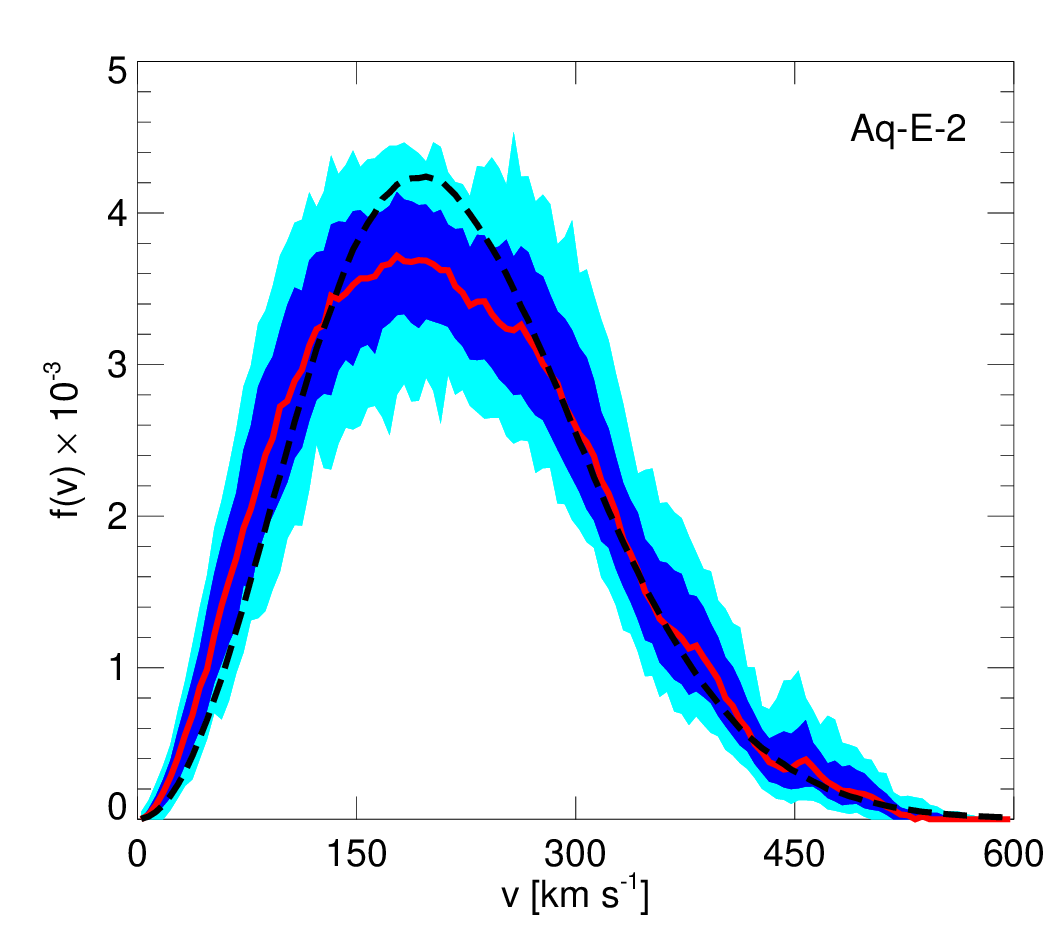}
\includegraphics[width=0.4\textwidth]{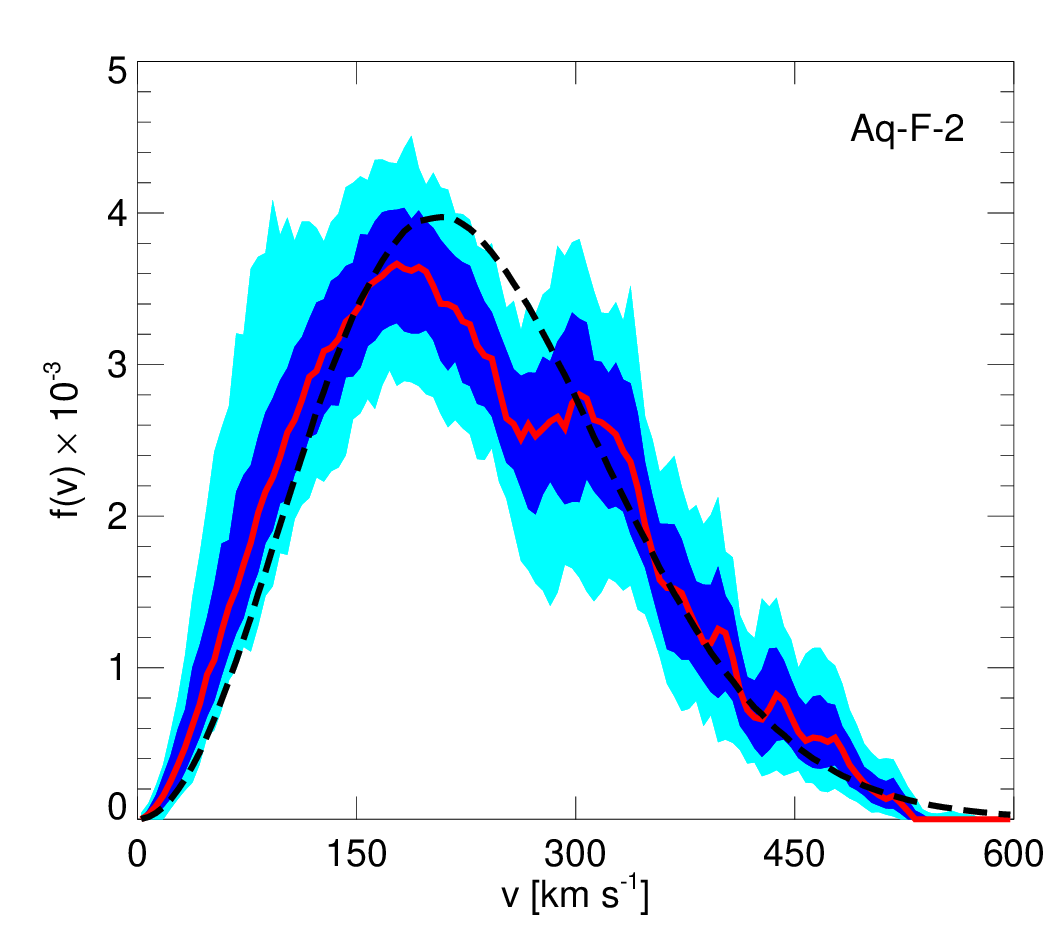}
}
\caption{Velocity modulus distributions in exactly the same format as
  the bottom panel of Fig.~\ref{fig:BoxVelA} but for all six of our
  halos at level-2 resolution. All distributions are smooth. Only in
  Aq-B-2 do we see a strong spike which is due to a single box which
  has 60\% of its mass (though a small fraction of its volume) in 
  a single subhalo.  No other box in any of
  the distributions has a subhalo contributing more than 1.5\% of the
  mass.  All distributions show characteristic broad bumps which are
  present in all boxes in a given halo, and so in its median
  distribution. These bumps are in different places in different
  halos.}
\label{fig:BoxVelAll} 
\end{figure*}
\begin{figure*}
\center{
\includegraphics[width=0.33\textwidth]{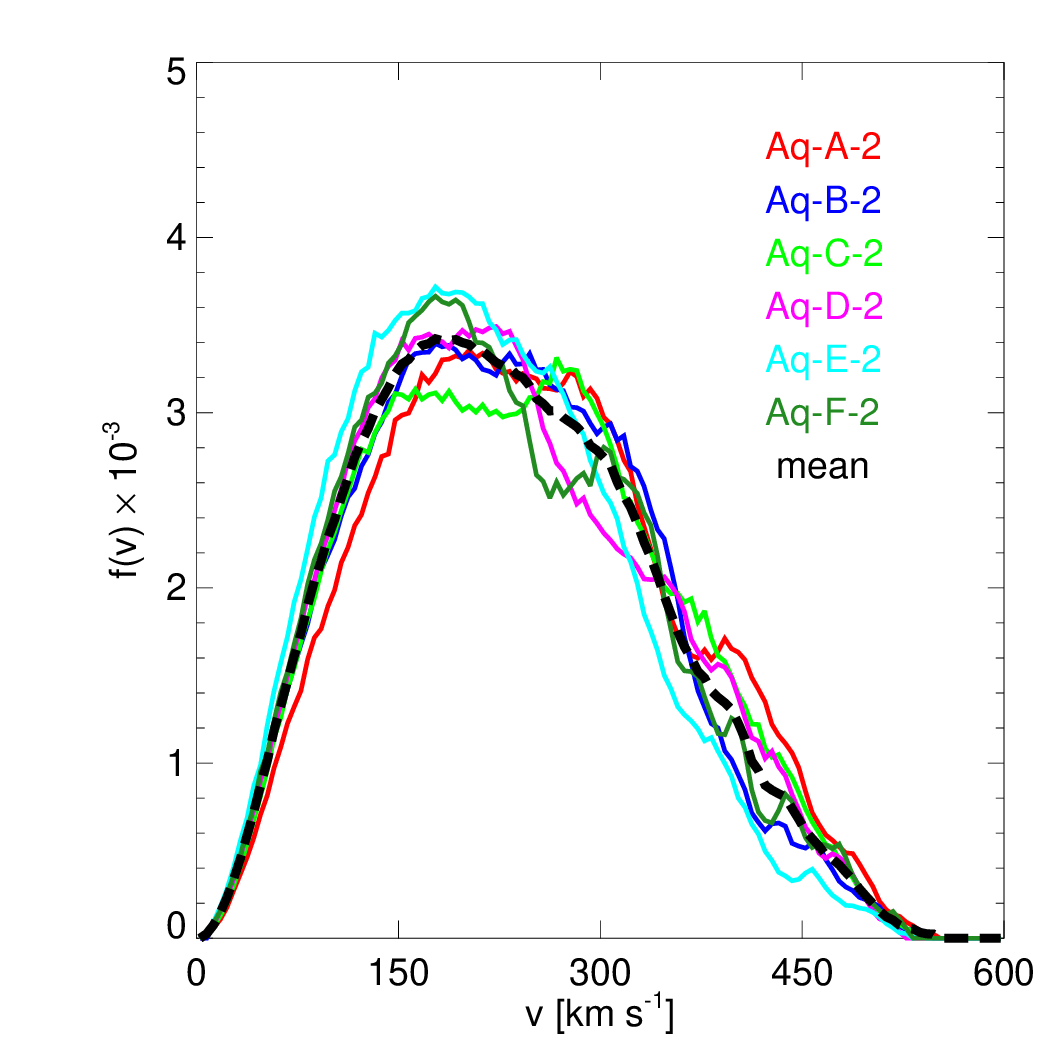}
\includegraphics[width=0.33\textwidth]{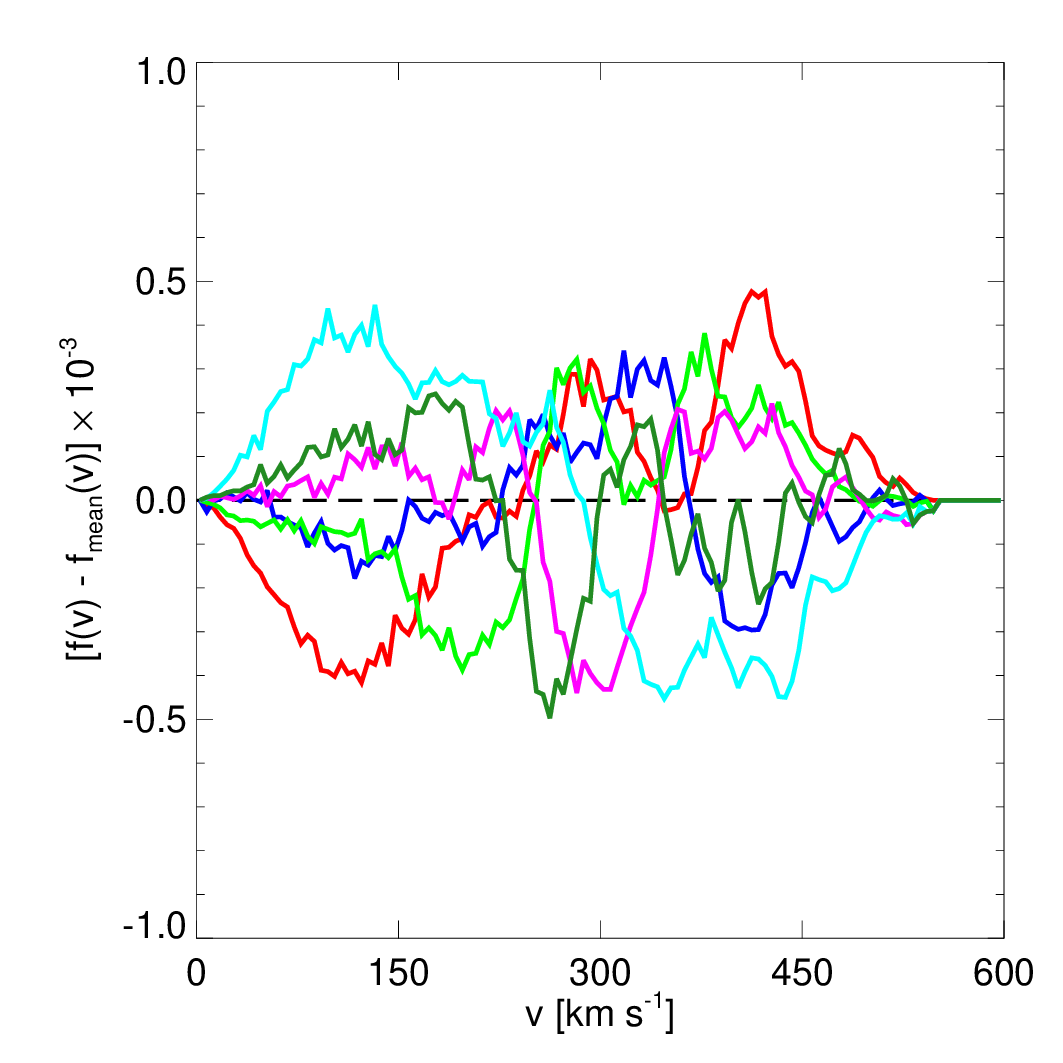}
\includegraphics[width=0.33\textwidth]{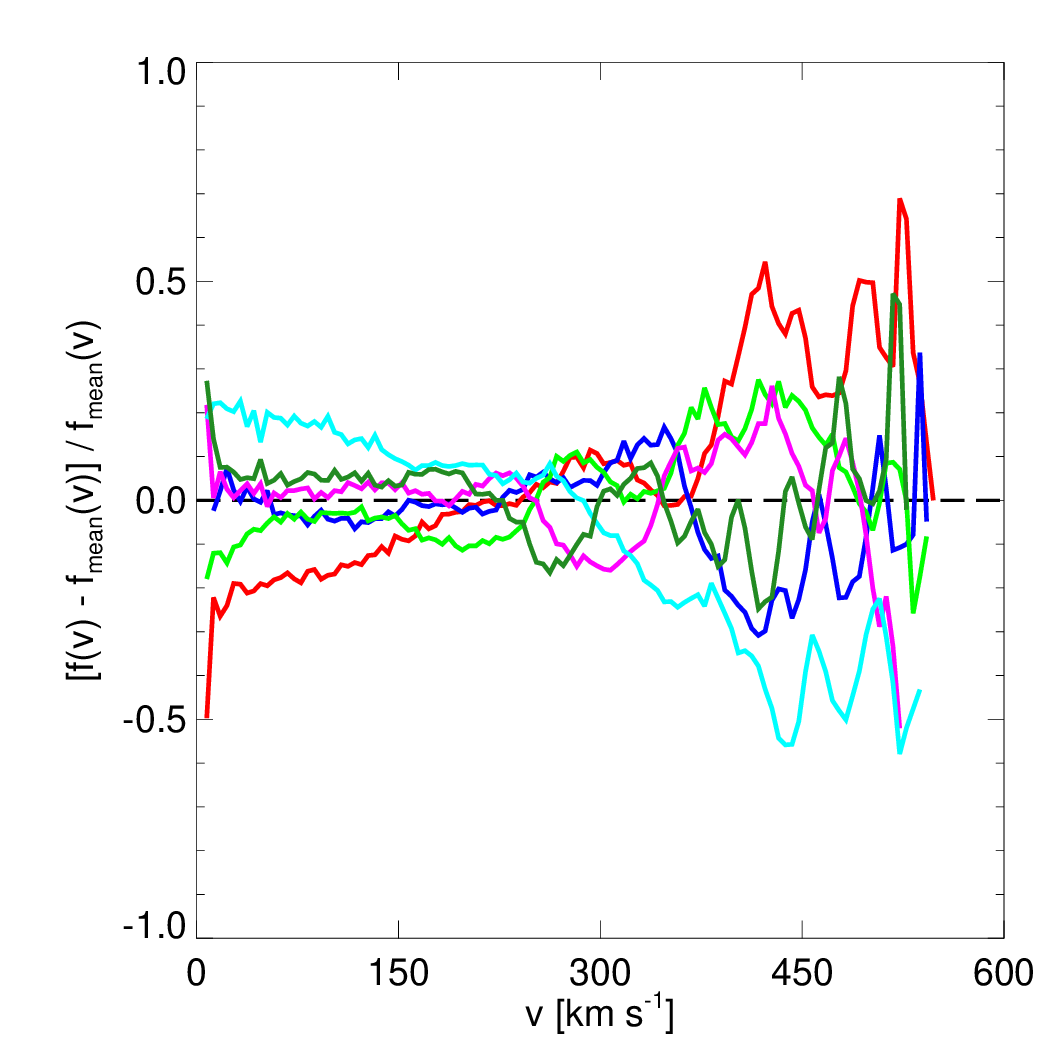}
}
\caption{Left panel: Median velocity modulus distributions for all six
  level-2 halos repeated from Fig.~\ref{fig:BoxVelAll}. The black
  dashed line is the mean of these distributions.  Middle panel:
  Deviations of the velocity modulus distribution of each of the six
  halos from the sample mean. The amplitude of the various bumps is
  similar in different halos and over the whole velocity range. It
  reaches more than 10\% of the amplitude of the mean
  distribution. Right panel: Relative deviations of the individual
  velocity modulus distributions from their sample mean. Typical
  relative deviations are about $30\%$, but they can exceed 50\% at
  higher velocities.  }
\label{fig:vel_mean} 
\end{figure*}
We address this issue by dividing the inner regions of each of our
halos into cubic boxes $2$~kpc on a side, and focusing on those boxes
centred between $7$ and $9$~kpc from halo centre. In Aq-A-1, each
$2$~kpc box contains $10^4$ to $10^5$ particles, while in the level-2
halos they contain an order of magnitude fewer. For every box we
calculate a velocity dispersion tensor and study the distribution of
the velocity components along its principal axes. In almost all boxes
these axes are closely aligned with those the ellipsoidal equidensity
contours discussed in the last section. We also study the distribution
of the modulus of the velocity vector within each box.  The upper four
panels of Fig.~\ref{fig:BoxVelA} show these distributions of a typical
$2$~kpc box at the solar circle in Aq-A-1 (solid red lines).  Here and
in the following plots we normalise distributions to have unit
integral. The black dashed lines in each panel show a multivariate
Gaussian distribution with the same mean and dispersion along each of
the principal axes. The difference between the two distributions in
each panel is plotted separately just above it, This particular box is
quite typical, in that we almost always find the velocity distribution
to be significantly anisotropic, with a major axis velocity
distribution which is platykurtic, and distributions of the other two
components which are leptokurtic.  Thus the velocity distribution
differs significantly from Maxwellian, or even from a multivariate
Gaussian. The individual velocity components have very smooth
distributions with no sign of spikes due to individual streams.  This
also is a feature which is common to almost all our 2~kpc boxes. It is
thus surprising that the distribution of the velocity modulus shows
clear features in the form of bumps and dips with amplitudes of
several tens of percent.

To see how these features vary with position, we overlaid the
distributions of the velocity modulus for all 2~kpc boxes centred
between 7 and 9~kpc from the centre of Aq-A-1 (bottom panel of
Fig.~\ref{fig:BoxVelA}). We superpose both the directly measured
distributions and the predictions from the best-fit multivariate
Gaussians. At each velocity, the solid red line show the median value
of all the directly measured distributions, while the dashed black
line is the median of all the multivariate Gaussian fits. The dark and
light regions enclose $68\%$ and $95\%$ of all the individual measured
distributions at each velocity.

It is interesting to note that the bumps in the velocity distribution
occur at approximately the same velocity in all boxes. This suggests
that they do not reflect local structures, but rather some global
property of the inner halo.  In
Fig.~\ref{fig:GaussianBoxVel_Resolution} we show velocity modulus
distribution for four different boxes in Aq-A at the three highest
resolutions (levels 1, 2 and 3). The error bars are based on Poisson
statistics in each velocity bin. Clearly the same bumps are present in
all boxes and at all resolutions. Thus, they are a consequence of real
dynamical structure that converges with increasing numerical
resolution.

\begin{figure}
\center{
\includegraphics[width=0.4\textwidth]{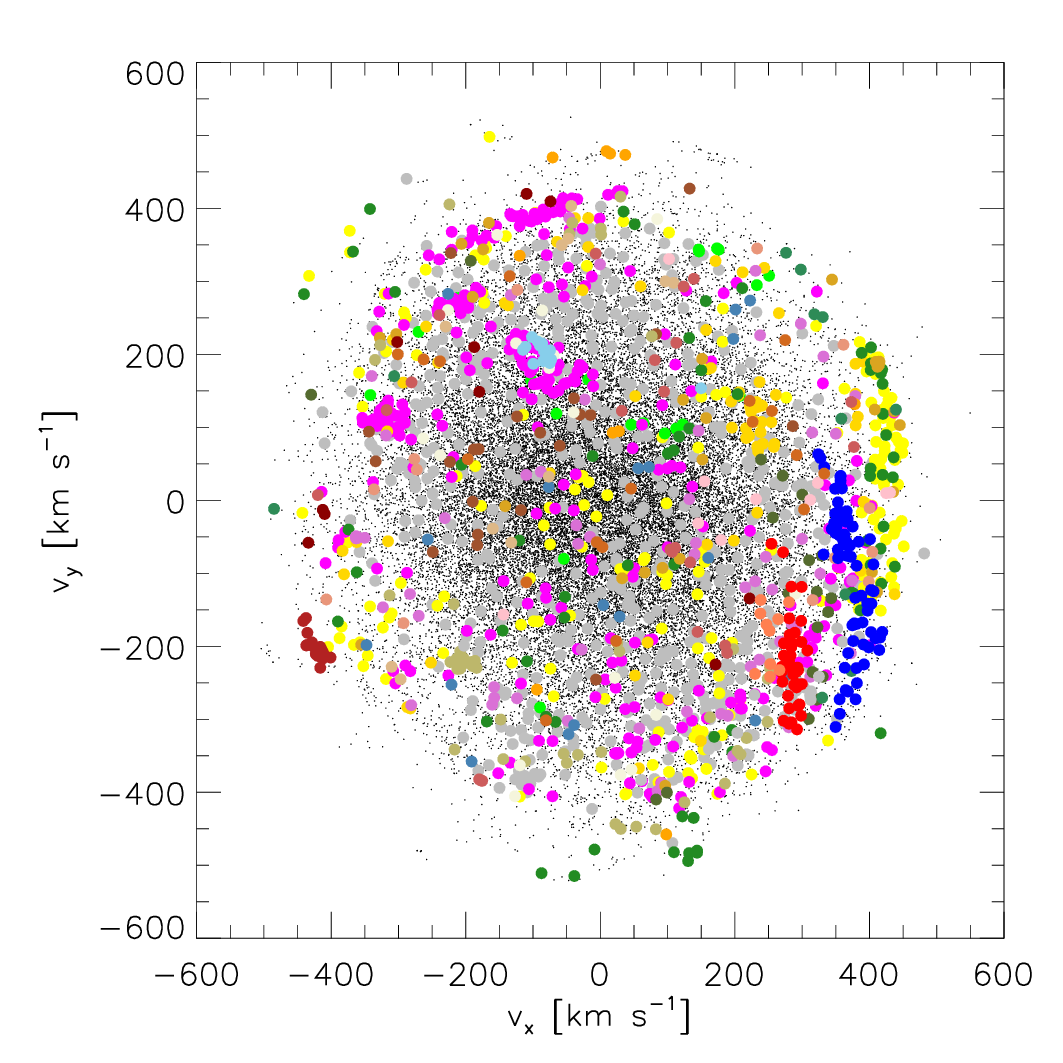}
\includegraphics[width=0.4\textwidth]{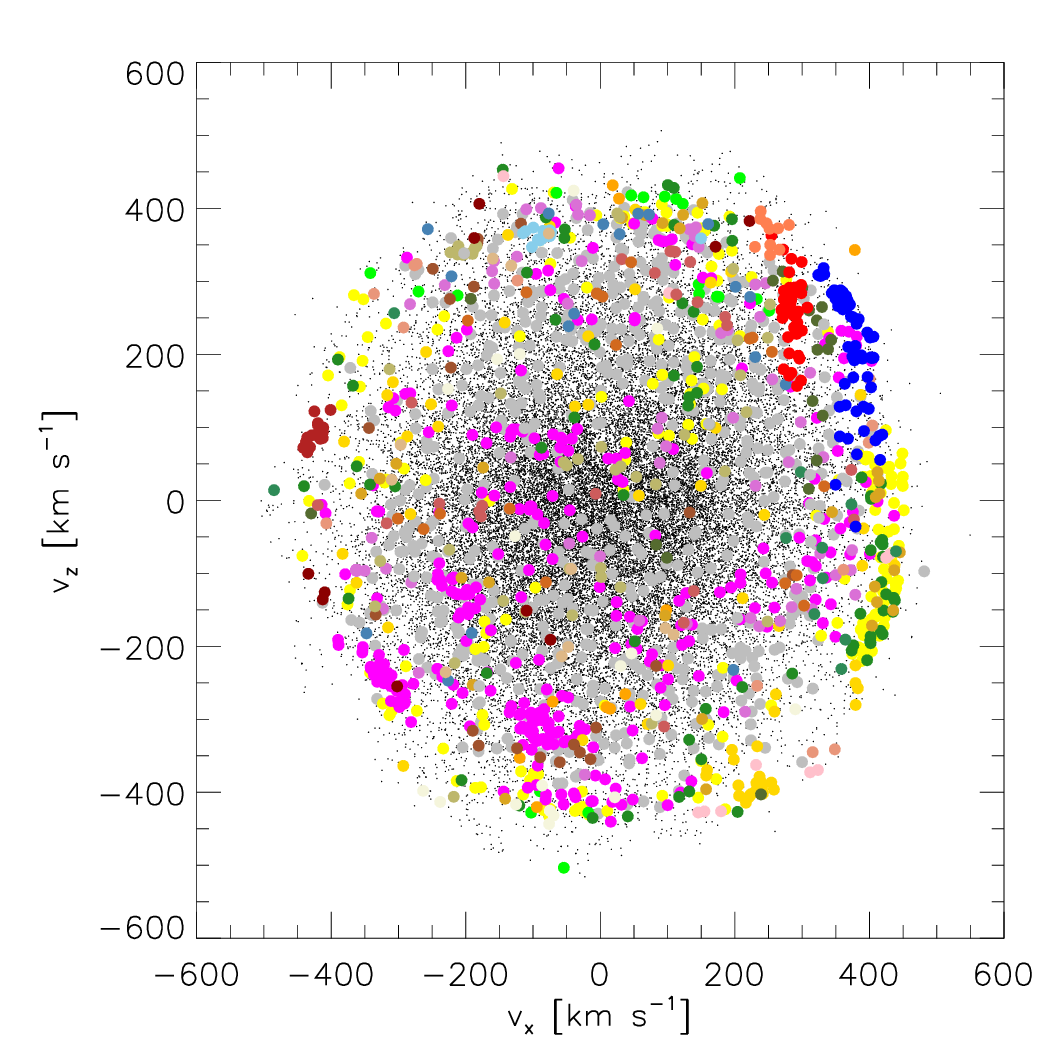}
\includegraphics[width=0.4\textwidth]{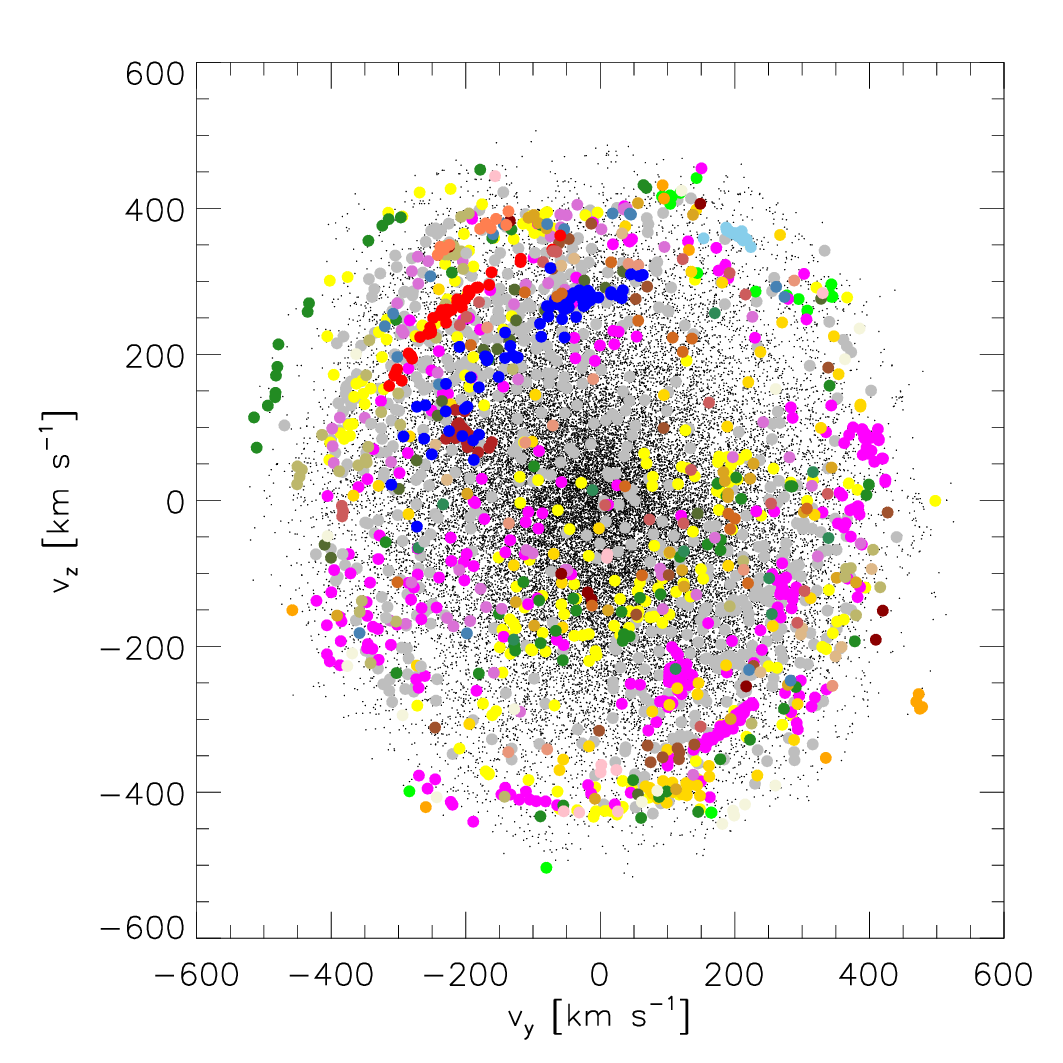}
}
\caption{Streams in velocity space for a $2$~kpc box $\sim 8$~kpc from
  the centre of Aq-A-1. Different colours stand for particles associated to
  different FoF groups at redshift $4.2$. Only groups contributing
  more than ten particles are shown. The box contains 27 such objects
  and has in total 41143 particles (shown as small black points)
  of which 1796 come from these groups. Clearly, particles originating 
  from the same group cluster in velocity space and build streams; often 
  many streams per group. }
\label{fig:streams} 
\end{figure}

In Fig.~\ref{fig:BoxVelAll} we make similar plots of the velocity
modulus distribution for all level-2 halos. These distributions are
quite smooth.  The sharp peak in Aq-B-2 is due to a single $2$~kpc box
where 60\% of the mass is contained in a single subhalo. No other box
in this or any other halo has more than $1.5$\% of its mass in a
single subhalo. The great majority of boxes contain no resolved
subhalo at all. Although the details of the median distributions vary
between halos, they share some common features.  The low velocity
region is more strongly populated in all cases than predicted by the
multivariate Gaussian model.  In all cases, the peak of the
distribution is depressed relative to the multivariate Gaussian. At
moderately high velocities there is typically an excess. Finally, and
perhaps most importantly, all the distributions show bumps and dips of
the kind discussed above. These features appear in different places in
different halos, but they appear at similar places for all boxes in a
given halo. The left panel of Fig.~\ref{fig:vel_mean} superposes the
median velocity modulus distributions of all level-2 halos and plots
their mean as a black dashed line. The middle panel shows the
deviations of the individual halos from this mean. The amplitudes of
the deviations are similar in different halos and at low and high
velocities. In percentage terms the deviations are largest at high
velocity reaching values of 50\% or more, as can be seen from the
right panel of Fig.~\ref{fig:vel_mean}.

The bumps in the velocity distribution are too broad to be explained
by single streams. Furthermore, single streams are not massive
enough to account for these features. This is shown more clearly in
Fig.~\ref{fig:streams} where we illustrate some streams in velocity space for
a $2$~kpc box in halo Aq-A-1. Different colours here indicate
particles that belonged to different FoF groups at redshift $4.2$. For
clarity we only show streams from groups that contribute at
least 10 particles to this volume ($0.025\%$ of the total number of
particles present at this location). There are 27 such objects.
If we consider all FoF groups that contribute more than $2$ particles
to the volume shown in Fig.~\ref{fig:streams}, we find that a given
FoF group contributes streams that are typically only populated by
$~2$ particles ($0.005\%$ of the total mass in the box). 
This implies that most of the groups contribute 
several streams of very low density. The most prominent streams have 
$\sim 40$ particles, i.e. $\sim 0.1\%$ of the mass in this volume. This clearly shows 
that streams are expected to be neither dense nor massive in the Solar vicinity.

The most prominent streams typically occupy the tail of the velocity
distribution in these local boxes. The excess of particles moving with
similar velocities and high kinetic energies can be measured using a
velocity correlation function, as shown by \cite{2002PhRvD..66f3502H}. 

\section{Energy distributions}

We have seen that the distributions of individual velocity components in
localised regions of space are very smooth, whereas the velocity
modulus distribution shows clear bumps. Taken together with the fact
that these bumps occur at similar velocities in regions on opposite
sides of the halo centre, this indicates that we must be seeing features
in the energy distribution of dark matter particles.
\begin{figure}
\center{
\includegraphics[width=0.4\textwidth]{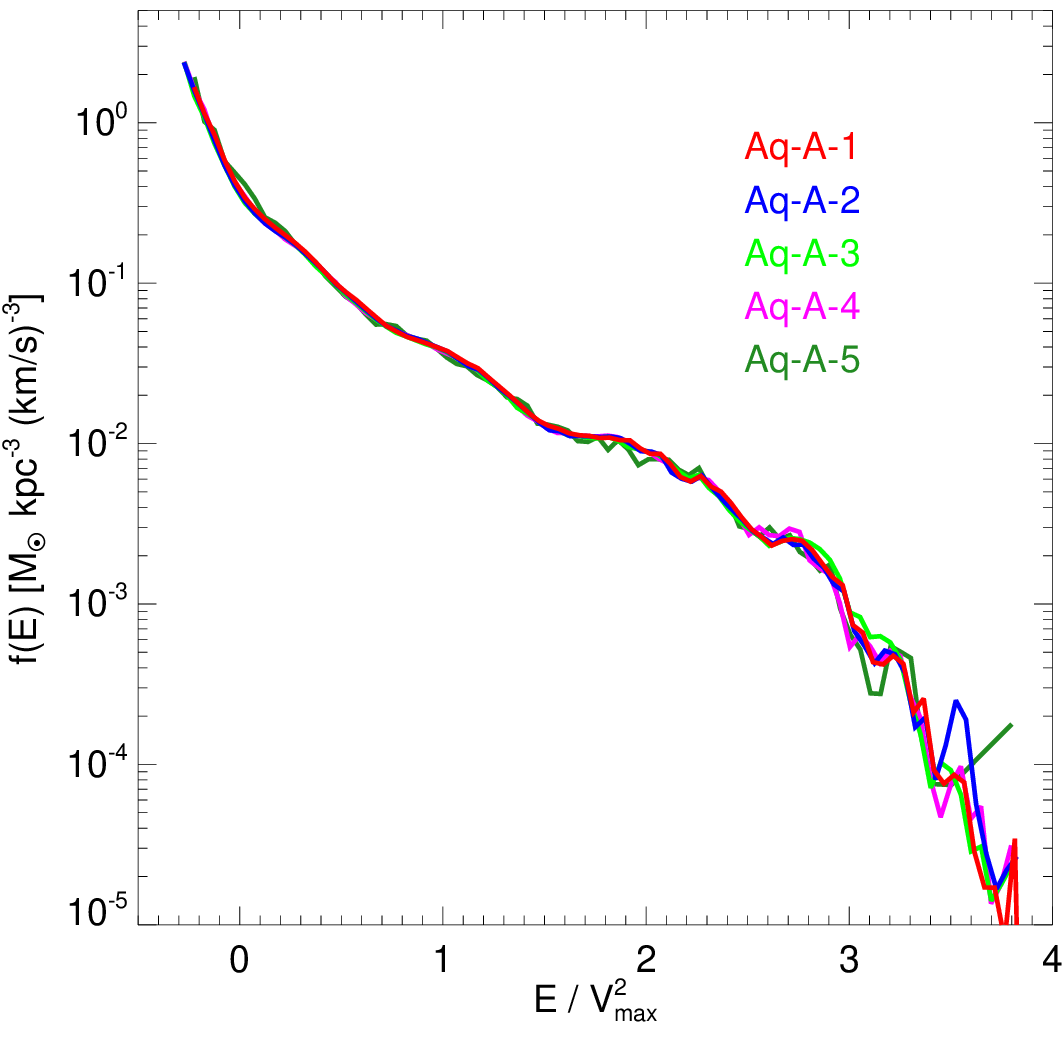}
}
\caption{Mean phase-space density distribution as a function of energy
  for Aq-A for particles in a spherical shell between $6$ and $12$~kpc
  and for all five resolution levels.  Especially at high-binding
  energies the convergence is very good. Features in the distribution
  function are visible at all resolutions for energies below
  $2.7~V_{\rm max}^2$, despite the fact that the mass resolution
  differs by more than a factor of $1800$ between Aq-A-1 and
  Aq-A-5. The less bound parts show more variation from resolution to
  resolution but still agree well between Aq-A-1 and Aq-A-2. }
\label{fig:DF_A_convergence} 
\end{figure}

To investigate this further, we estimate the mean phase-space density
as a function of energy in each of our halos using the properties
of the particles at radii between $6$ and $12$~kpc.  Clearly our halos
are not perfectly in equilibrium and they are far from spherical. Thus
their phase-space densities will only approximately be describable
as functions of the integrals of motion, and they will depend
significantly on integrals other than the energy. Nevertheless
we can estimate a mean phase-space density as a function of
energy by taking the total mass of particles with $6~{\rm kpc}< r
<12~{\rm kpc}$ and energies in some small interval and dividing it by
the total phase-space volume corresponding to this radius and energy
range, e.g. 
\begin{equation}
f(E) = \frac{\mathrm{d}M}{\mathrm{d}E}~\frac{1}{g(E)},
\end{equation}
where $f(E)$ is the energy-dependent mean phase-space density and
\begin{equation}
g(E) \,\, = 
\, 4 \pi \!\!\!\!\!\!\! 
\int\limits_{\mathcal{V},E>\Phi({\bf x})} \!\!\!\!\!\!\!\! \mathrm{d}^3 {\bf x} \,\,\, \sqrt{2\,(E-\Phi(\bf x))},
\end{equation}
is the available phase-space volume in the configuration-space volume 
$\mathcal{V}$. The differential energy
distribution is easily calculated by binning the energies of all
particles between $6$ and $12$~kpc. The phase-space volume can be
calculated by solving for the gravitational potential at the position
of all simulation particles and then using these as a Monte-Carlo
sampling of configuration space in the relevant integrals.  Taking the
ratio then yields the desired estimate of $f(E)$.

In Fig.~\ref{fig:DF_A_convergence} we show $f(E)$ measured in this way
for all our simulations of Aq-A. We express the energy in units of
$V_{\rm max}^2$ and we take the zero-point of the gravitational potential
to be its average value on a sphere of radius 8~kpc. As a result the
measured energy distribution extends to slightly negative values. Note
how well the distribution converges at the more strongly bound
energies. At higher energies the convergence between the level-1
and 2 resolutions is still very good. This demonstrates that we can robustly
measure the mean phase-space density distribution. Furthermore, we see
clear wiggles that reproduce quite precisely between the different
resolutions.

\begin{figure}
\center{
\includegraphics[width=0.4\textwidth]{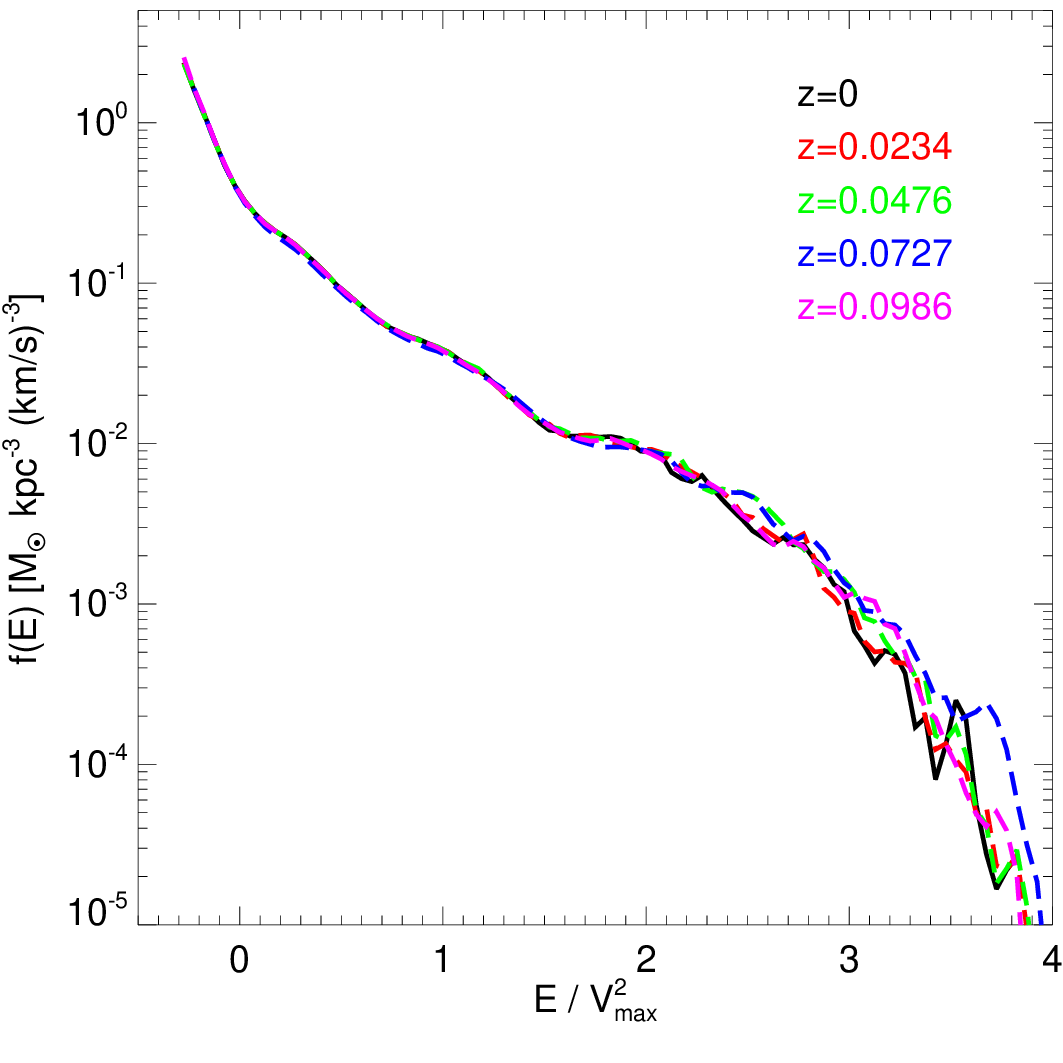}
}
\caption{Evolution of the mean phase-space density distribution of
  Aq-A-2 over four time intervals of about $300$ Myr. Below $2.4~V_{\rm max}^2$ 
  the phase-space distribution function is time-independent,
  implying that the system has reached coarse-grained equilibrium. The
  small bumps at these energies are therefore well-mixed features in
  action space. The variability of the features in the weakly bound
  part of the distribution shows that they are due to individual
  streams and therefore change on the timescale of an orbital period.
  Note that the phase-space density at these energies is almost three
  orders of magnitude below that of the most bound particles.  }
\label{fig:DF_A_redshift} 
\end{figure}
\begin{figure}
\center{
\includegraphics[width=0.4\textwidth]{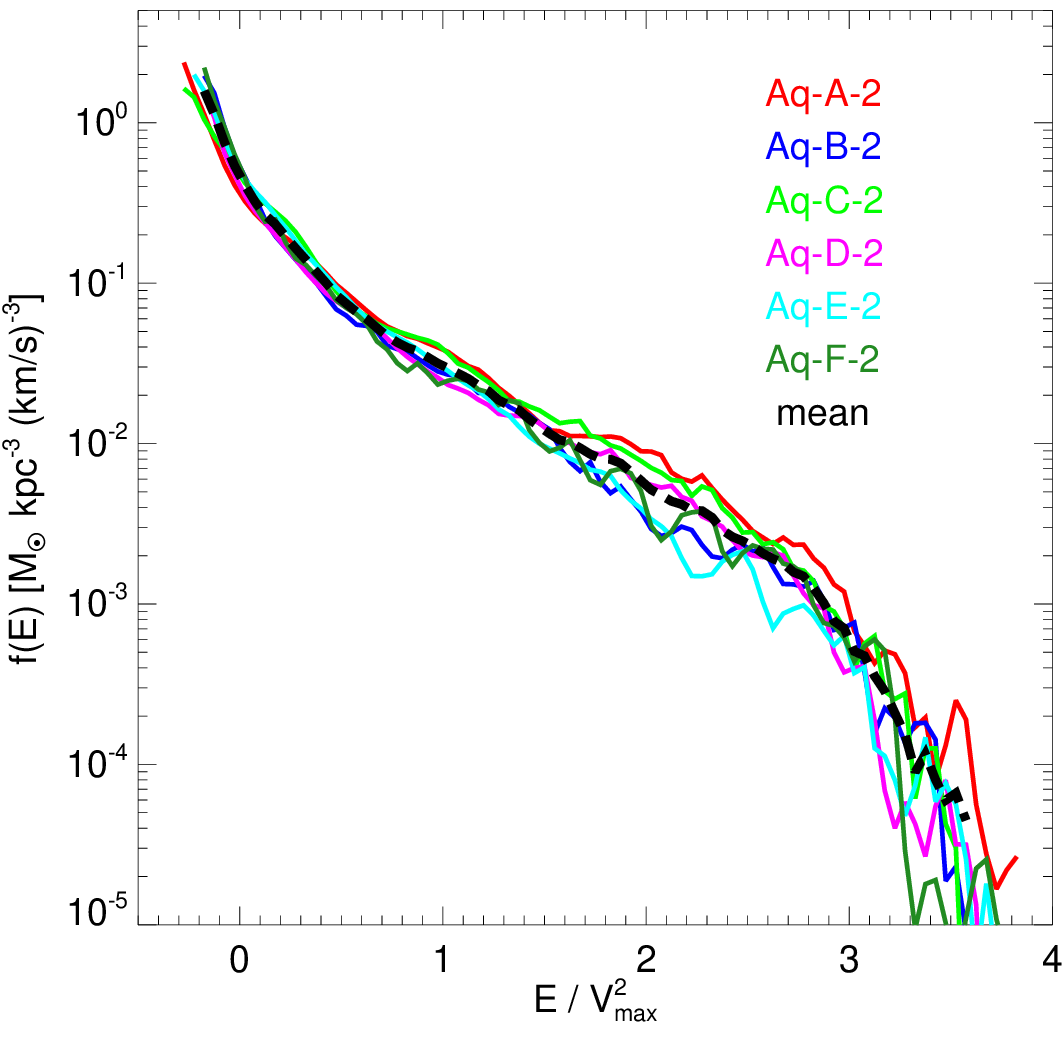}
}
\caption{Scaled phase-space distribution functions for all level-2
  halos. In addition to scaling according to $V_{\rm max}$ we have
  also corrected for a zero-point offset in the potential energy
  between different halos. The black dashed line shows the average
  distribution function based for our halo sample. At high binding
  energies the scatter between average and individual halo
  distribution functions is quite small, showing that this part of the
  distribution function is near-universal. At low-binding energies
  large amplitude features are visible in all halos. These features
  differ from halo to halo and are related to recent events in their
  formation histories.  }
\label{fig:DF_2_scaled} 
\end{figure}

Fig.~\ref{fig:DF_A_redshift} shows similarly estimated mean
phase-space density distributions for Aq-A-2 at five different times
separated by about $300$ Myr. This is longer than typical orbital
periods in the region we are studying.  Despite this, the wiggles at
energies below $2.4~V_{\rm max}^2$ are present over the complete
redshift range shown.  This demonstrates that these features are
well-mixed, and that the phase-space distribution function has reached
a coarse-grained equilibrium. In contrast, the variability of the
wiggles in the part of the distribution corresponding to weakly bound
particles (where the orbital periods are much larger) shows that these
must be due to individual streams or to superpositions of small
numbers of streams, which have not yet phase-mixed away.

\begin{figure}
\center{
\includegraphics[width=0.4\textwidth]{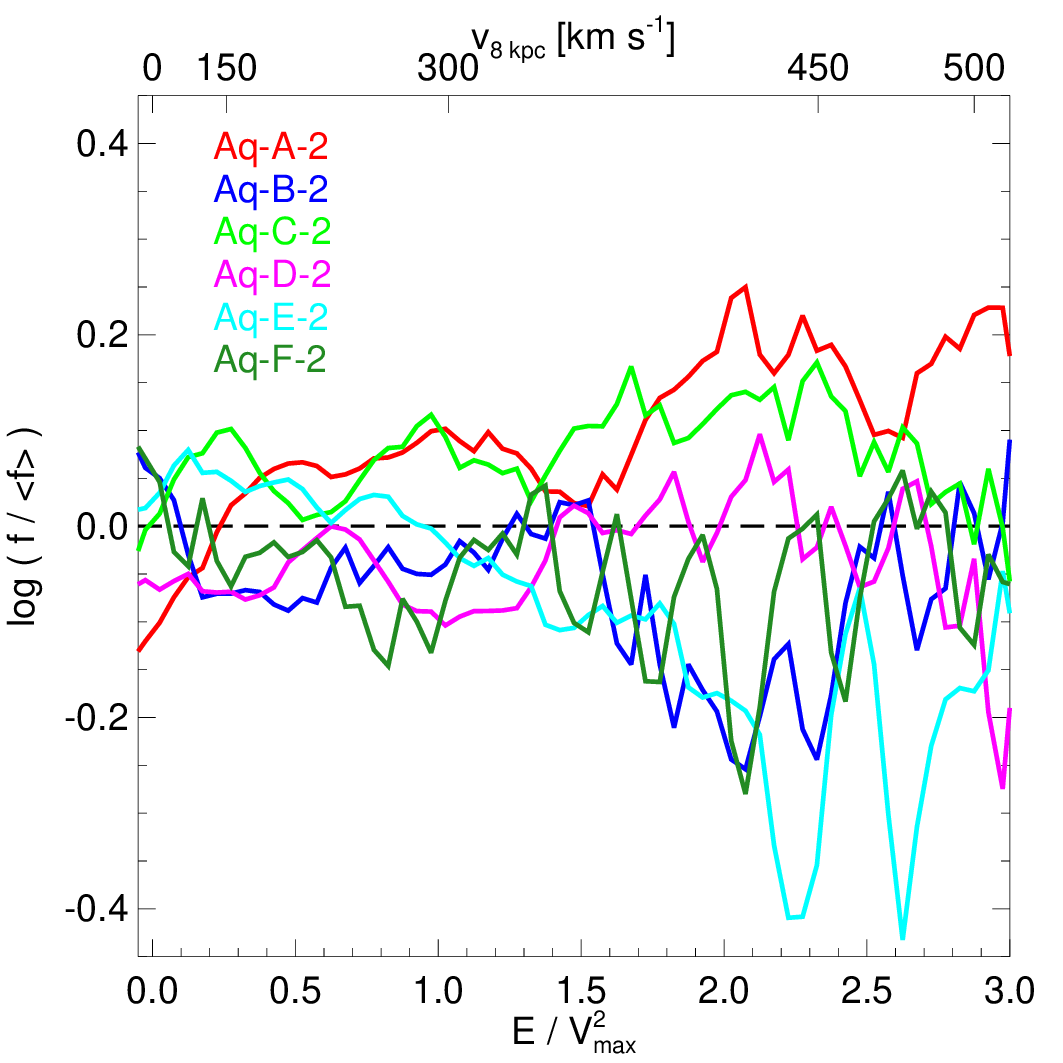}
}
\caption{Deviations of the individual phase-space density
  distributions from the mean over our sample of level-2 halos. We
  focus here on the more bound part. The lower $x$-axis shows the
  orbital energy while the upper one shows the corresponding velocity
  $8$~kpc distance from halo centre.  The amplitude of features
  increases for $V_{8~{\rm kpc}}>350$~km/s.  At even lower binding
  energies, $E>3~V_{\rm max}$ deviations can reach an order of
  magnitude, see Fig.~\ref{fig:DF_2_scaled}. }
\label{fig:DF_2_scaled_scatter} 
\end{figure}
\begin{figure*}
\center{
\includegraphics[width=0.4\textwidth]{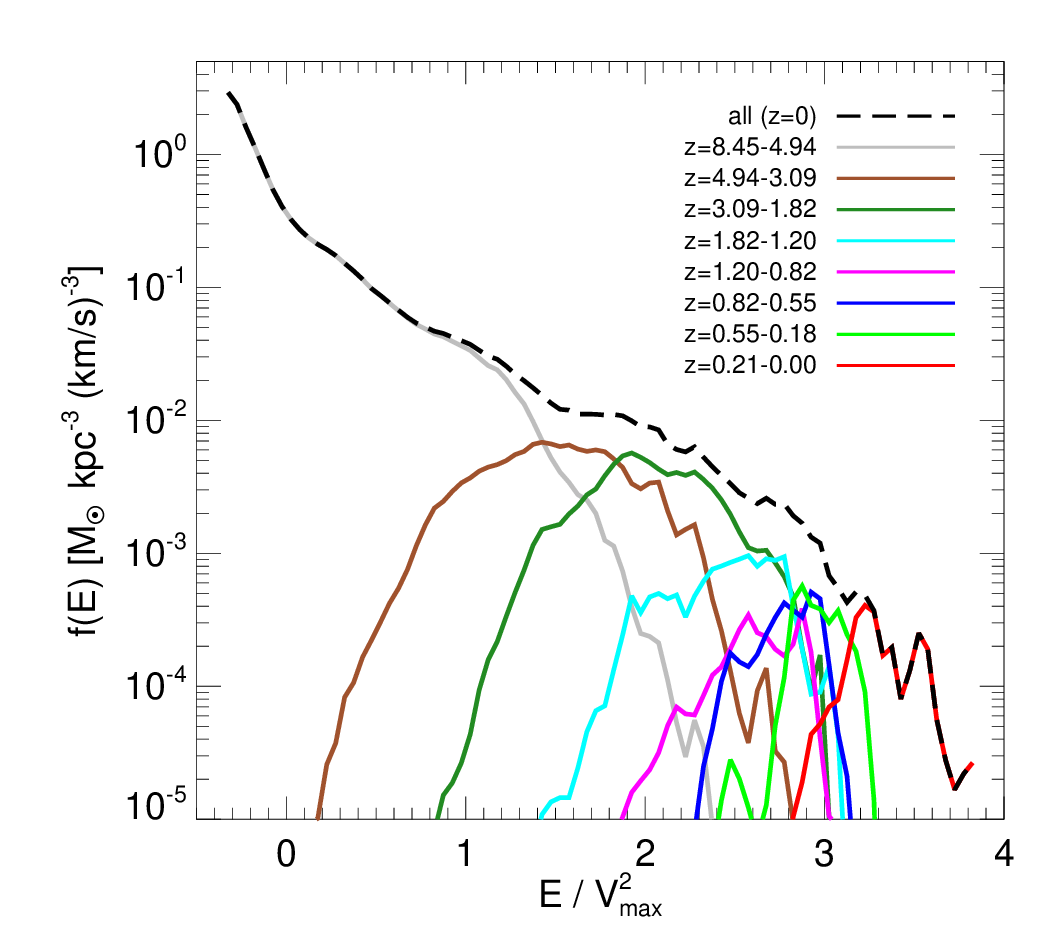}
\includegraphics[width=0.4\textwidth]{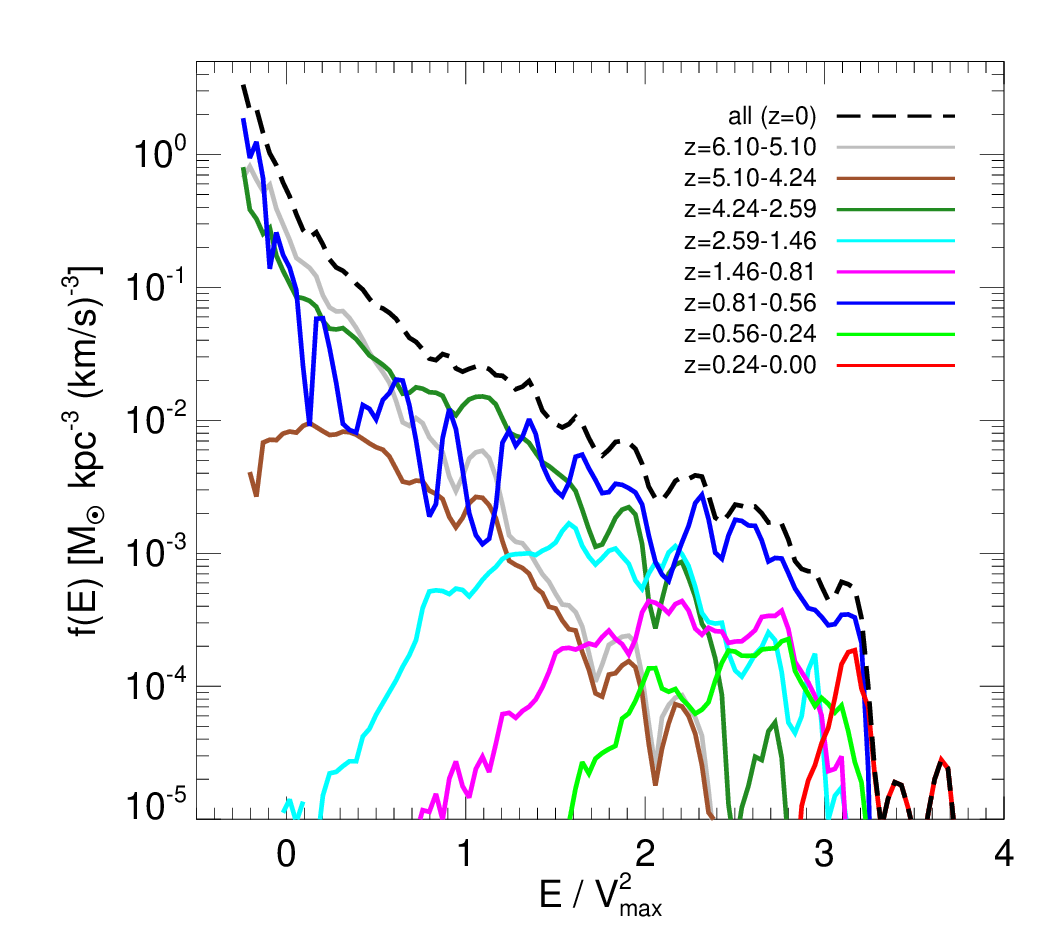}
\includegraphics[width=0.4\textwidth]{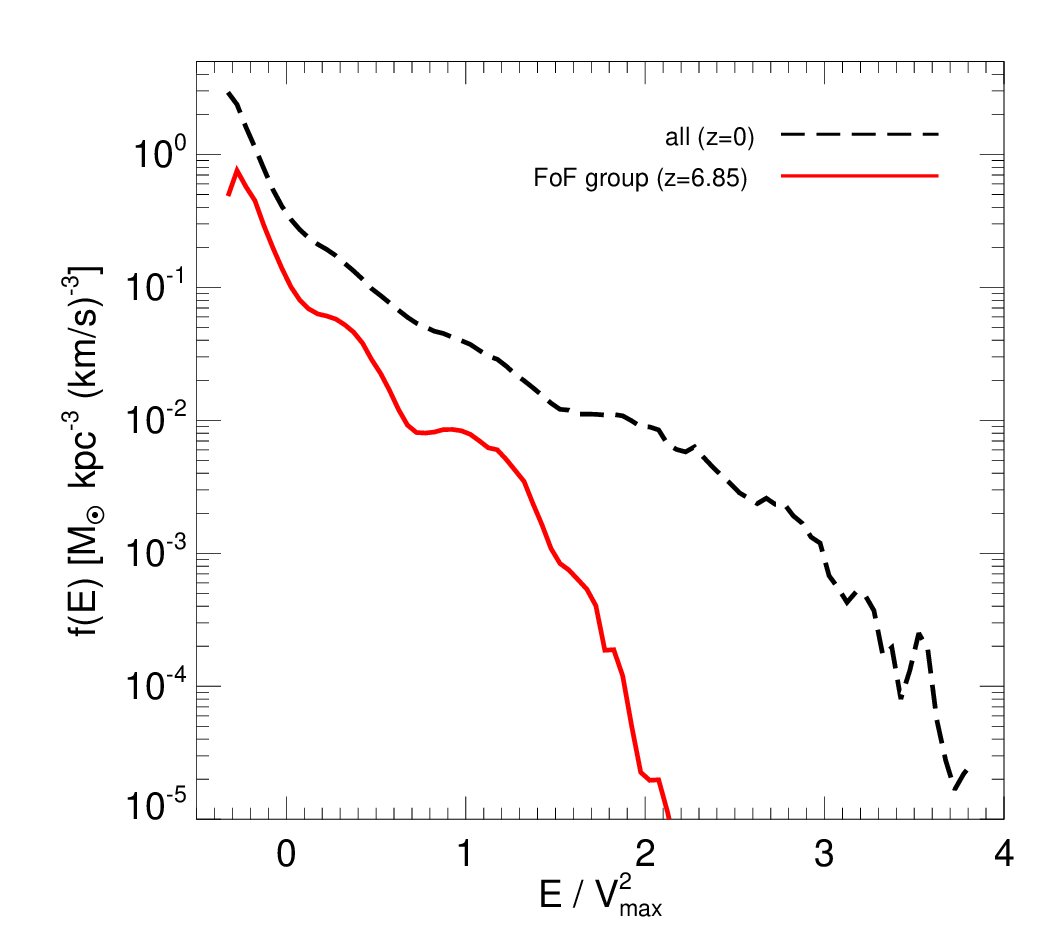}
\includegraphics[width=0.4\textwidth]{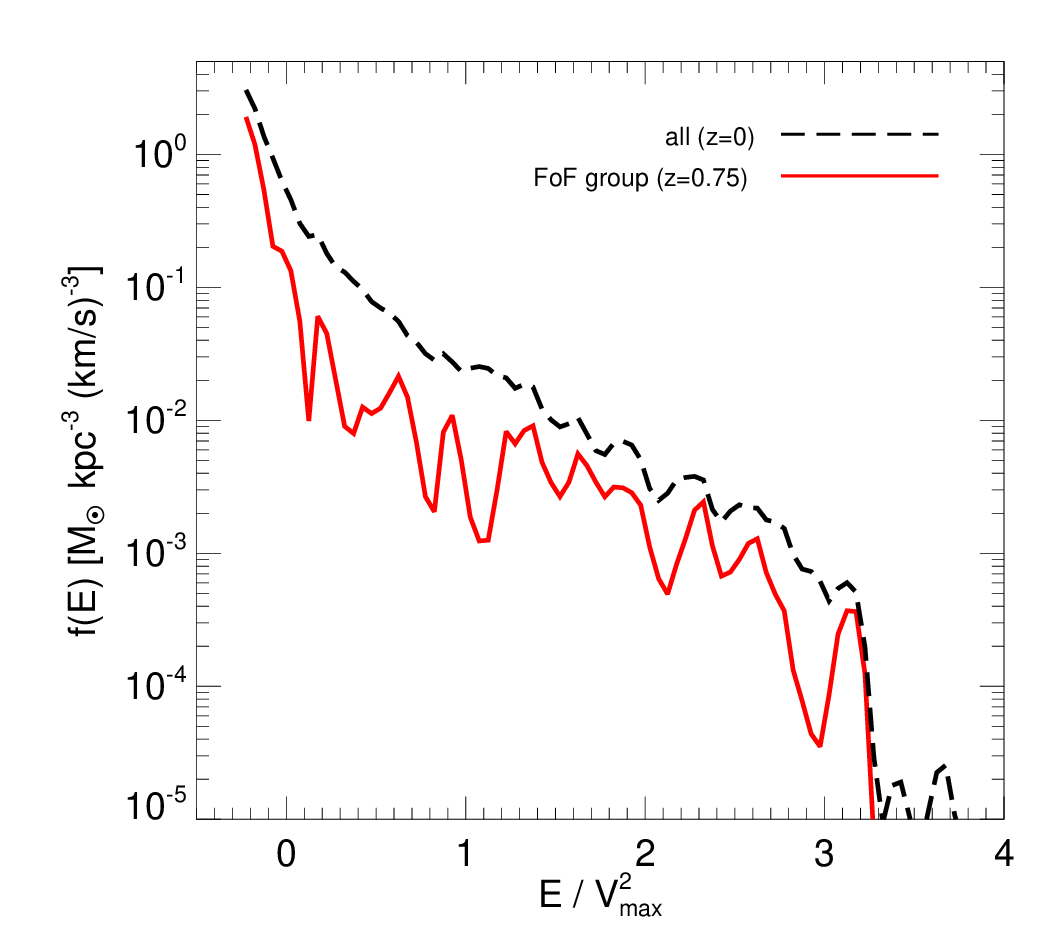}
}
\caption{Top row: Contributions to the present-day phase-space density
  distribution from particles accreted at different epochs (indicated
  by different colours). The top left panel shows the build-up of the
  distribution function for halo Aq-A-2. This halo has a quiescent
  formation history with no recent mergers.  The top right panel is a
  similar plot for Aq-F-2, which underwent a major merger between $z=0.75$ and $z=0.68$. 
  Bottom row: These plots isolate the contribution of a
  single, massive FoF group to the $z=0$ phase-space density
  distributions. For Aq-A-2 this group was identified at $z=6.85$, for
  Aq-F-2 at $z=0.75$. In both cases it is clear that material from
  the group is responsible for some of the features seen in the
  present-day phase-space density distribution.}
\label{fig:DF_Formation} 
\end{figure*}

To estimate what these phase-space distribution functions should look
like for a ``smooth'' system, we average the functions found in our
six individual level-2 halos. In Fig.~\ref{fig:DF_2_scaled} we
superpose these six functions and their mean $\langle f \rangle$ (the
black dashed line).  The similarity of the different distribution
functions at high binding energies suggests a near-universal shape for
$f(E)$. At lower binding energies, individual halos deviate quite
strongly from $\langle f \rangle$.  This can be seen more clearly in
Fig.~\ref{fig:DF_2_scaled_scatter} where we plot $\log(f/\langle f
\rangle)$, the decimal logarithm of the ratio of the phase-space
density of an individual halo to the mean. The lower axis is orbital
energy in units of $V_{\rm max}^2$, while the upper axis is the
corresponding dark matter particle velocity at the Solar Circle. In
this plot one can clearly see the wiggles, which are located at
different energies for different halos. For $V_{8~{\rm kpc}}<
350$~km/s the distribution functions for all halos satisfy $0.7 <
f/\langle f \rangle < 1.4 $. For low binding energies (velocities of
600~km/s or more at the Solar Circle) this ratio can exceed a factor
of ten.

These features in the phase-space density distribution must be related
to events in the formation of each halo. To demonstrate this
explicitly, we have computed $f(E)$ separately for particles which
were accreted onto two of our halos (i.e. first entered the main
progenitor FoF group) at different epochs. The upper left panel of
Fig.~\ref{fig:DF_Formation} shows that Aq-A-2 had a very ``quiet''
merger history. Material accreted at different times is arranged in a
very orderly way in energy space. All the most strongly bound
particles were accreted before redshift 5, and material accreted at
successively later times forms a series of ``shells'' in energy
space. The most weakly bound wiggles are due entirely to the most
recently accreted material, and progressively more bound bumps can be
identified with material accreted at earlier and earlier times.  In
contrast, the top right panel shows that Aq-F-2 had a very ``active''
merger history, with a major merger between $z=0.75$ and $z=0.68$. The correspondence
between binding energy and epoch of accretion is much less regular
than for Aq-A-2, and much of the most bound material actually comes
from the object which fell in between $z=0.75$ and $z=0.68$. It is also striking that
many of the wiggles in this object are present in material that
accreted at quite different times, suggesting that they may be
non-steady coherent oscillations rather than stable structures in
energy space.  Nevertheless, in both halos one can identify features
in the phase-space density distribution with particles accreted at
certain epochs, and in both halos the most weakly bound particles were
added only very recently. Note that the phase-space density of this
material is very low, so it contributes negligibly to the overall
local dark matter density. In the bottom panels of
Fig.~\ref{fig:DF_Formation} we show the $f(E)$ distributions of
particles which were associated with a single, massive FoF-group which
was identified at $z=6.85$ in the case of Aq-A-2 and at $z=0.75$ in the
case of Aq-F-2.  The wiggles in the strongly bound part of Aq-A-2 are
clearly due to this early merger event, while the later merger in
Aq-F-2 is responsible for most of the material accreted in
$0.56<z<0.81$ and for most of the strong features in
the phase-space density distribution.

\begin{figure*}
\center{
\includegraphics[width=0.4\textwidth]{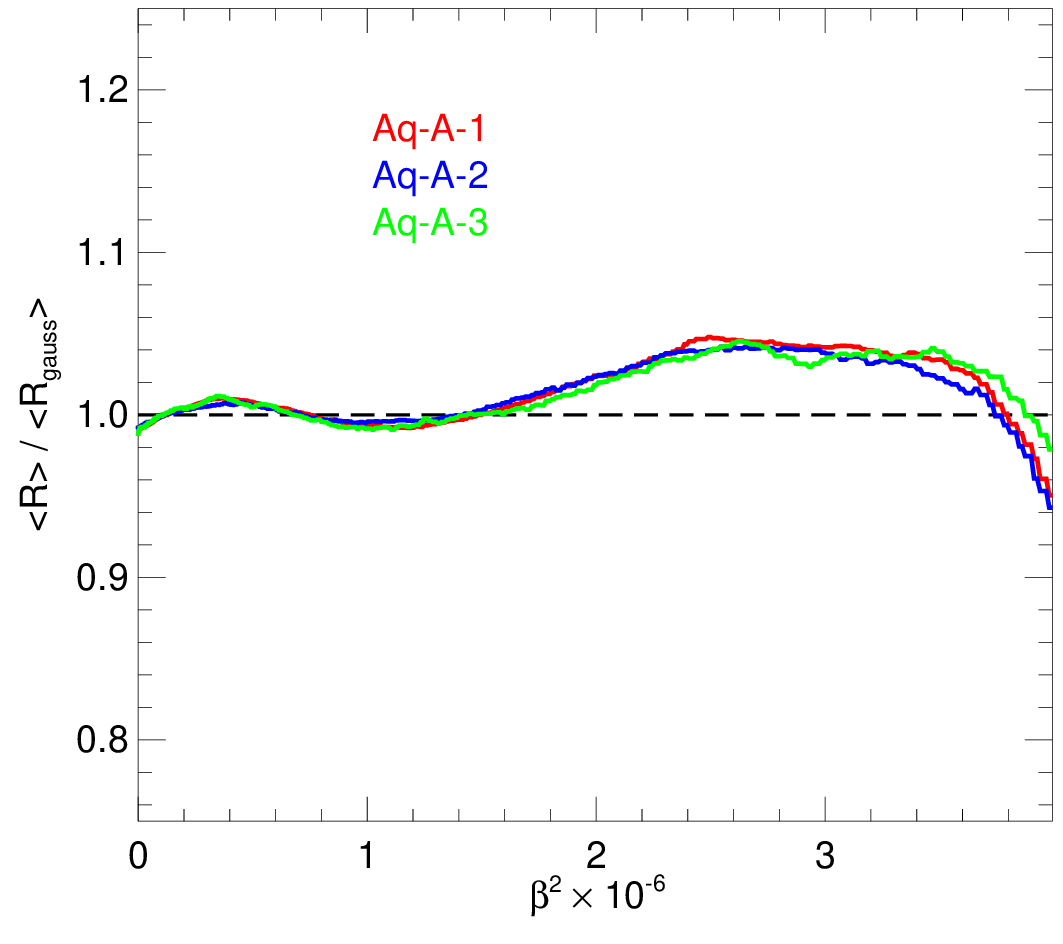}
\includegraphics[width=0.4\textwidth]{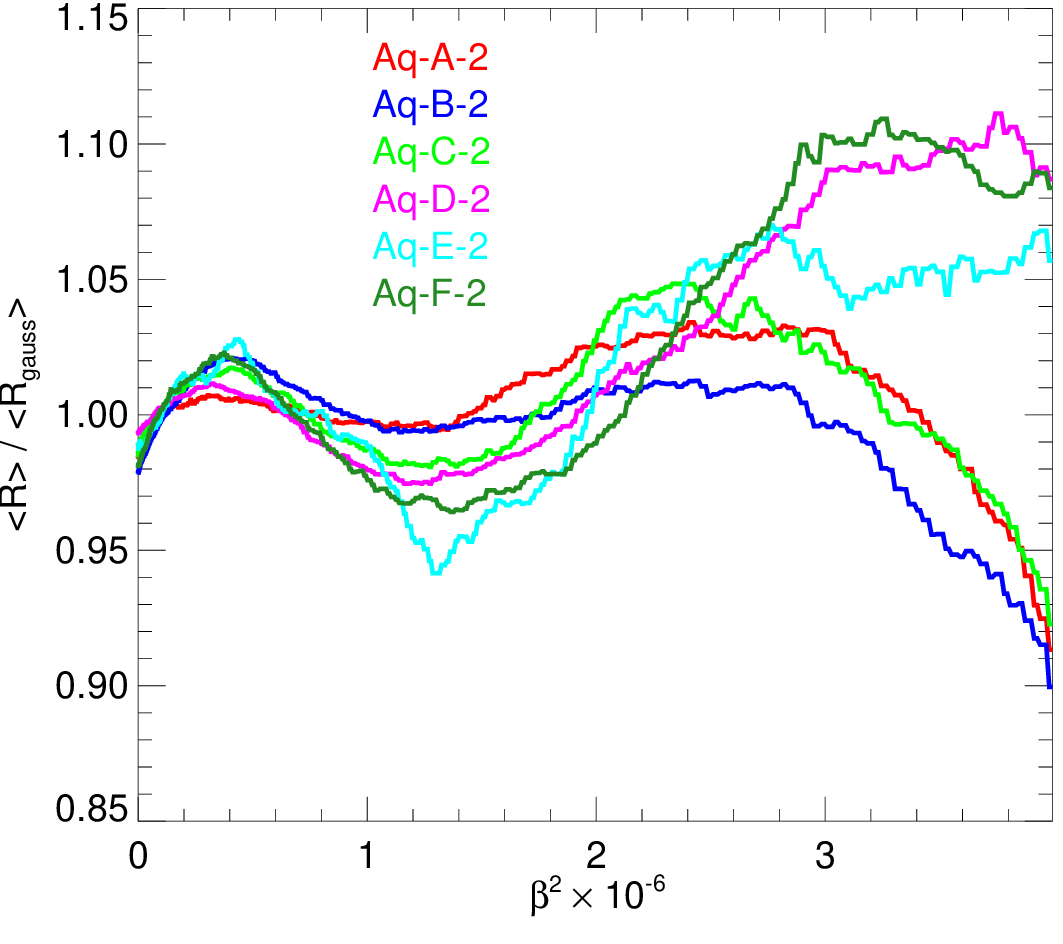}
}
\caption{
  Recoil spectra ratio for the three highest resolution simulations of Aq-A (left) and 
  the level-2 (right) simulations of the other halos. For these plots 
  we averaged the recoil rate over a year for every box and then calculated 
  the median recoil rate ratio $\langle R \rangle$/$\langle R_{\rm gauss} \rangle$ 
  of the rates for the simulation and for the best-fit multivariate Gaussian 
  distribution. The $x$-axis is directly  proportional to the energy.  
  In all level-2 halos the expected recoil spectrum 
  based on a multivariate Gaussian can be wrong by about $10\%$ depending on the 
  energy. Furthermore the behaviour of the deviations seems quite similar.
  This is due to the fact that the velocity distributions differ in a 
  characteristic way from a multivariate Gaussian. The deviations in the 
  recoil spectra are typically highest at high energies.
}
\label{fig:WIMP_spectrum} 
\end{figure*}

We conclude that these features in the energy distribution should open
the window to ``dark matter astronomy'' once experiments reach the
sensitivity needed for routine detection of DM particles. We will then
be able to explore the formation history of the Milky Way using the DM
energy distribution.

\section{Detector signals}

We will now use the spatial and velocity distributions explored
above to calculate expected detector signals. The main
question here is how the non-Gaussian features of the velocity
distribution influence these signals. Our results show that
features due to subhalos or massive streams are expected to be
unimportant. On the other hand, deviations of the velocity distributions from a perfect 
Gaussians in terms of general shape, bumps and dips can 
have an impact on detector signals.

There are currently more than 20 direct detection experiments
searching for Galactic DM, most of them focusing on WIMPs.  For these,
the detection scheme is based on nuclear recoil with the
detector material. The differential WIMP elastic scattering rate can
be written as \citep{1996PhR...267..195J}:
\begin{equation}
R = \mathcal{R} ~ \rho_0 ~ T(E,t), 
\end{equation}
where $\mathcal{R}$ encapsulates the particle physics parameters 
(mass and cross-section of the WIMP; form factor and mass of target nucleus),
$\rho_0$ is the local dark matter density that we assume to be constant 
based on the results of section 3, and
\begin{equation}
T(E,t) = \int\limits_{v_{\rm min}}^\infty \!\!\! \rm{d} v \,\,\,\frac{f_v(t)}{v} , 
\end{equation}
where $f_v$ is the WIMP speed distribution in the rest frame of the
detector integrated over the angular distribution. $v_{\rm min}$ here is
the detector-dependent minimum WIMP speed that can cause a recoil of
energy $E$:
\begin{equation}
v_{\rm min} = \left ( \frac{E~(m_\chi + m_A)^2}{2 m_\chi^2 m_A} \right )^{1/2},
\end{equation}
where $m_\chi$ is the WIMP mass and $m_A$ the atomic mass of the target
nucleus. To get detector independent results we set $\mathcal{R}=1$
in the following\footnote{This also implies that we assume the form factor
to be constant. Any other form factor will change the shape of the recoil
spectrum. Since we are not interested in the exact shape of the spectrum,
but in deviations expected due to different velocity distributions, we
neglect form factor effects in the following.}. 

\begin{figure*}
\center{
\includegraphics[width=0.4\textwidth]{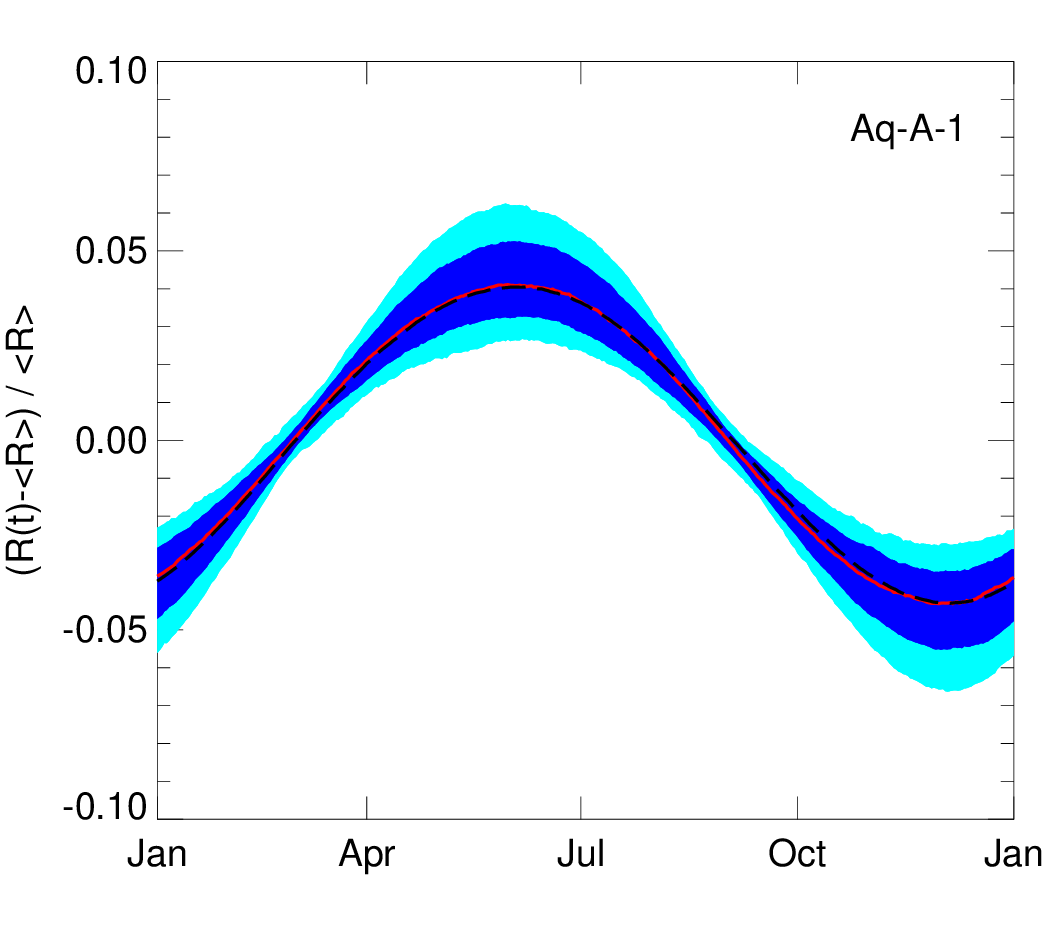}
\includegraphics[width=0.4\textwidth]{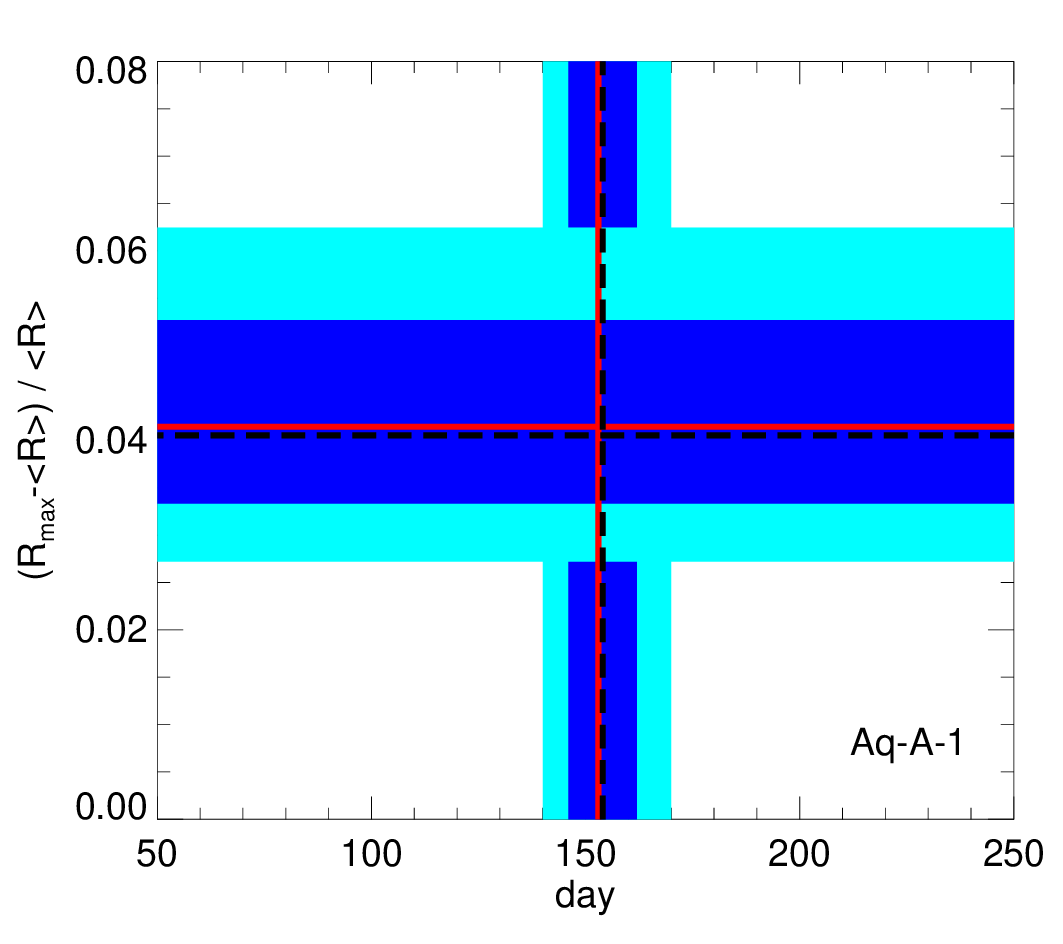}
\includegraphics[width=0.4\textwidth]{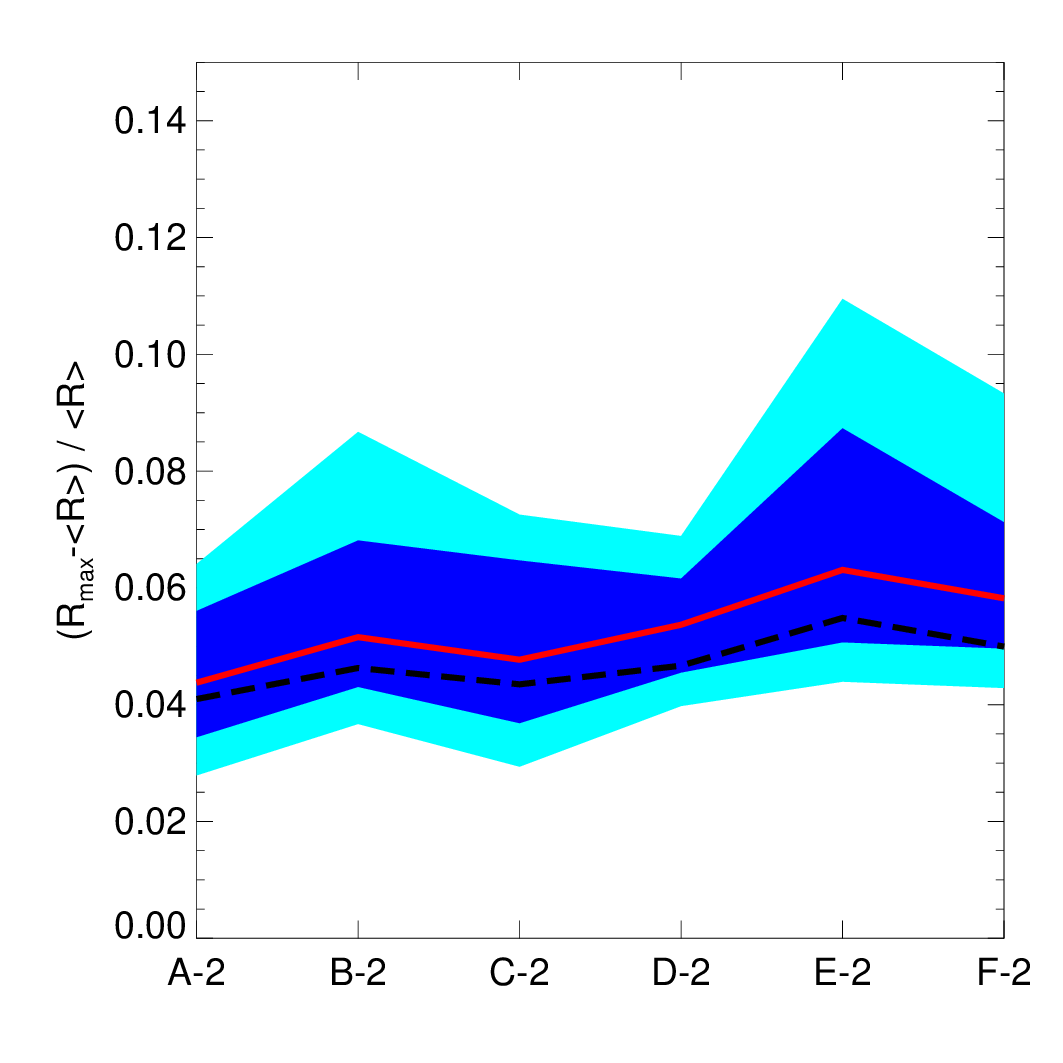}
\includegraphics[width=0.4\textwidth]{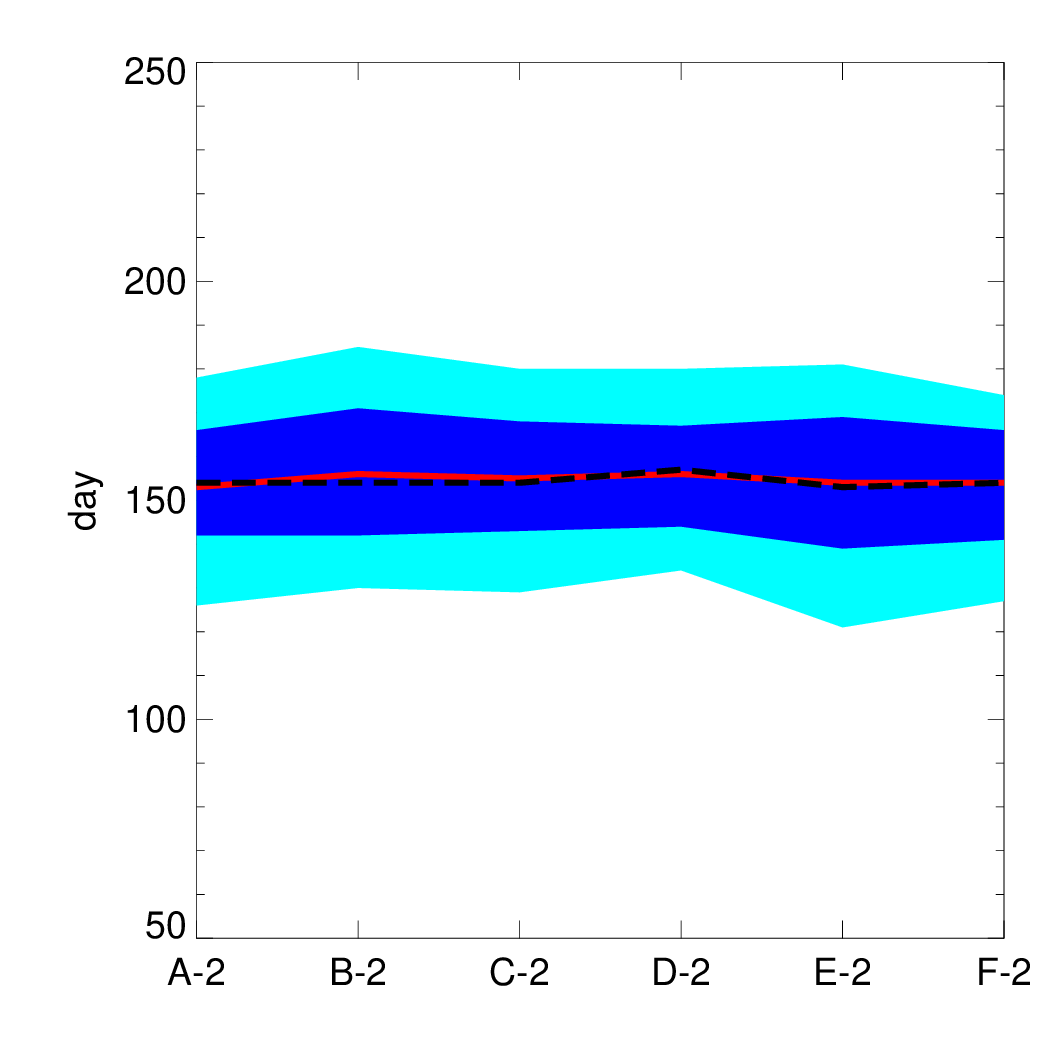}
}
\caption{Top panels: Annual modulation for all $2$~kpc boxes with  
  halocentric distance between $7$ and $9$~kpc in halo Aq-A-1 assuming
  $v_{\rm min}=300$~km/s.  The left plot shows how the dimensionless recoil
  rate $(R(t)-\langle R \rangle )/\langle R \rangle $ changes over the
  year. The right plot shows the corresponding modulation parameter
  space defined by the peak day ($x$-axis) and maximum amplitude
  $(R_{\rm max}-\langle R \rangle )/\langle R \rangle $ ($y$-axis).
  Bottom panels: Modulation parameters for the local $2$~kpc
  boxes of all level-2 resolution halos. There is no clear trend visible 
  in the day of maximum behaviour over the halo sample. On the other hand, the 
  median amplitude in all boxes is higher than expected based on the 
  Gaussian sample for $v_{\rm min}=300$~km/s. The line and contour 
  scheme is the same as in Fig.~\ref{fig:BoxVelA}.}
\label{fig:ModulationAll} 
\end{figure*}

The recoil rate shows a annual modulation over the year
\citep{1986PhRvD..33.3495D}. To take this into account we add the Earth's
motion to the local box velocities to transform Galactic rest
frame velocities into the detector frame.  We model the motion of the
Earth according to \cite{1996APh.....6...87L} and \cite{1998gaas.book.....B}.
Let $\vec{v}_E=\vec{u}_r+\vec{u}_S+\vec{u}_E$ be the velocity of the
Earth relative to the Galactic rest frame decomposed into Galactic rotation 
$\vec{u}_r$, the Sun's peculiar motion $\vec{u}_S$ and the Earth's velocity 
relative to the Sun $\vec{u}_E$. In Galactic coordinates 
these velocities can be written as:
\begin{align}
\vec{u}_r &=(0,222.2,0) ~\rm{km/s}, & \nonumber \\ 
\vec{u}_S &=(10.0,5.2,7.2) ~\rm{km/s}, & \nonumber \\
e_{E,i} &=u_E(\lambda) ~\cos(\beta_i)  ~\sin(\lambda-\lambda_i), & \nonumber \\ 
u_E(\lambda) &= \langle u_E \rangle ~[1- e \sin(\lambda - \lambda_i)]&
\end{align}
where $i=R,\phi,z$, $\lambda$ is the ecliptic longitude
($\lambda_0=(13 \pm 1)^\circ$),$\langle u_E \rangle=29.79$~km/s is
the mean velocity of the Earth around the Sun, and the ellipticity of the Earth orbit
$e=0.016722$. The $\vec{u}_r$ value is based on a combination of a large number
of independent determinations of the circular velocity
by \cite{1986MNRAS.221.1023K}. We note that this value has 
a standard deviation of $20$~km/s. For the constant $\beta$ and $\lambda$ 
angles we take:
\begin{align}
(\beta_R,\beta_\phi,\beta_z) &= (-5.5303^\circ,59.575^\circ,29.812^\circ)& \nonumber \\
(\lambda_R,\lambda_\phi,\lambda_z) &= (266.141^\circ, -13.3485^\circ, 179.3212^\circ)
\end{align}
The ecliptic longitude can be written as
\begin{align}
\lambda(t) &= L(t) + 1.915^\circ \sin g(t) + 0.020^\circ \sin 2g(t), & \nonumber  \\
L(t) &=280.460^\circ + 0.9856474^\circ t,& \nonumber \\
g(t) &=357.528^\circ + 0.9856003^\circ t,& 
\end{align}
where $t$ is the fractional day number relative to noon (UT) on 31
December 1999 (J2000.0).  We refer to a day number relative to 31
December 2008 in our plots. In what follows we will assume that the
$R$-direction is always aligned with the major axis of the principal
axis frame of the velocity ellipsoid in each box and the $\phi$- and
$z$-directions, with the intermediate and short axes. This is needed to
add the Earth's motion to the box velocities, and to transform the
velocity vectors in each box to the detector frame.

Clearly the deviations of the velocity distribution from a perfect
multivariate Gaussian found in the previous sections will also alter 
the recoil spectrum, because the velocity integral $T(E,t)$ effectively 
measures the $1/v$-weighted area under the velocity curve. As in the previous 
sections we compare the results obtained directly from the simulations to the 
expectation for a best-fit multivariate Gaussian distribution.
In Fig.~\ref{fig:WIMP_spectrum} we plot recoil spectra ratios for the three highest
resolution simulations of Aq-A (left) and the level-2 (right) simulations of the other 
halos. For these plots we averaged the recoil rate over a year for individual boxes.
The rates are calculated using the simulation velocity distribution
($\langle R \rangle$) and the best-fit Gaussian model for each box 
($\langle R_{\rm gauss} \rangle$).  The plots show the median of the ratios
$\langle R \rangle / \langle R_{\rm gauss} \rangle$ over all boxes.  Since we assume 
that the density $\rho_0$ is constant in a given box, it drops out when calculating 
the ratios. The $x$-axis  measures the energy in dimensionless $\beta=v/c$ 
values. For a given detector this can easily be converted to keV, assuming the 
masses $m_\chi$ and $m_A$ are given in GeV/$c^2$:
\begin{equation}
E=\frac{2~m_\chi^2~m_A}{(m_\chi+m_A)^2}~c^2~~\beta^2~~\times~~10^6~~{\rm keV}
\end{equation}

Fig.~\ref{fig:WIMP_spectrum} clearly shows that in all level-2 halos the
expected recoil spectrum based on a multivariate Gaussian model can differ
by up to $10\%$ from the directly predicted simulation result. Furthermore, the behaviour of
the deviations seems to be similar in all cases, especially at low energies,
where we already found that the phase-space density is nearly universal.
The similarity in the deviations between the different halos is due to the 
fact that the velocity distributions all differ in a characteristic way from
the Gaussian distributions as shown in section 4. The deviations 
in the recoil spectra are typically highest at high energies.

\begin{figure}
\center{
\includegraphics[width=0.4\textwidth]{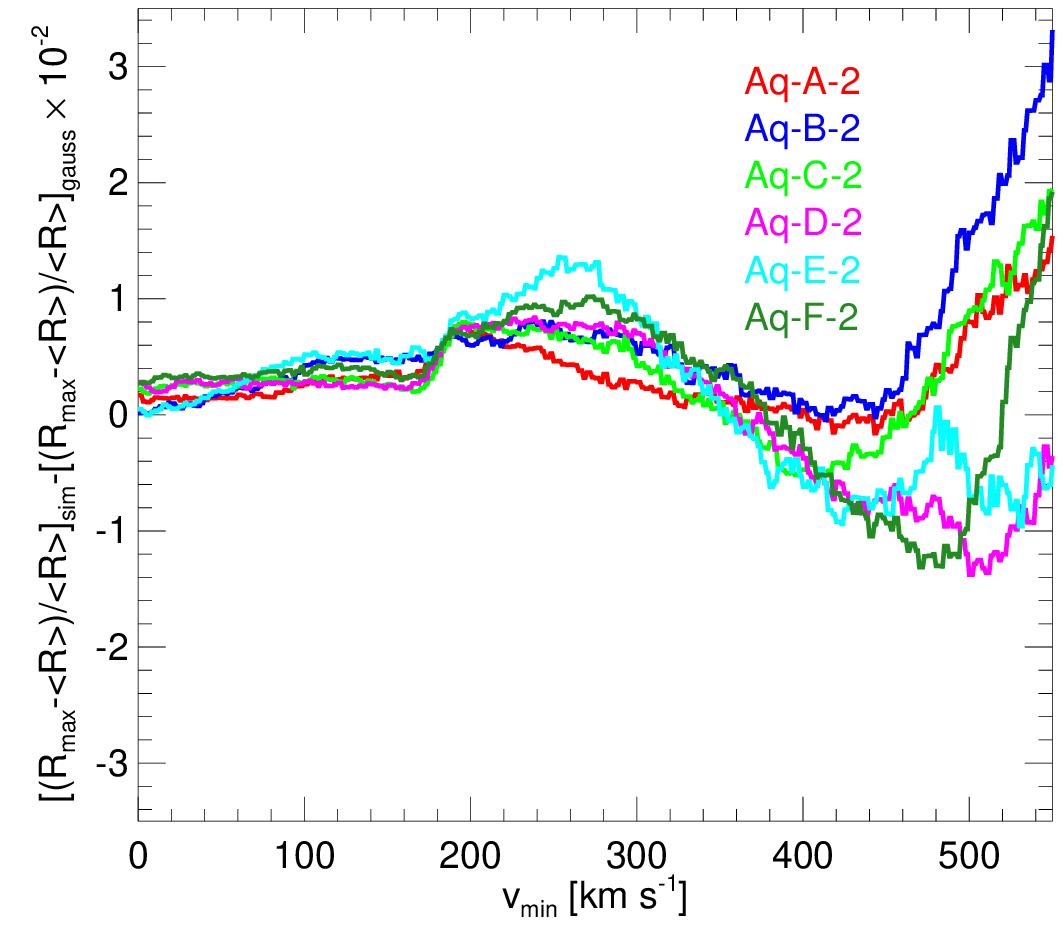}
\includegraphics[width=0.4\textwidth]{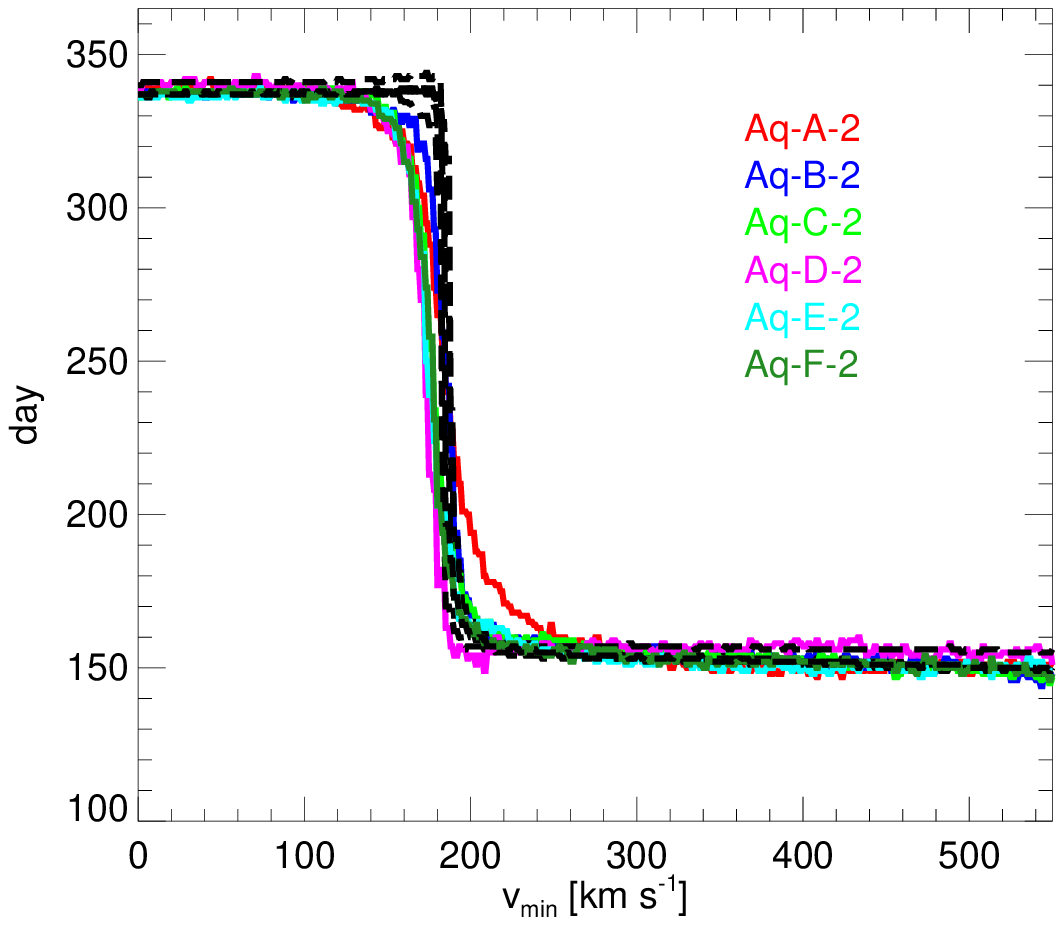}

}
\caption{Top panel: Median recoil rate amplitude for all $2$~kpc
  boxes with halocentric distance between $7$ and $9$~kpc for
  all level-2 halos. The plot shows the difference between the 
  relative maximum modulation amplitude observed in the simulation and that expected 
  for the best-fit multivariate Gaussian distribution. Bottom panel: Median day of maximum
  amplitude for the same halos (solid red) compared to their Gaussian
  predictions (dashed black). The day of maximum amplitude is the
  same for all boxes and is well reproduced in the Gaussian model. The
  phase-reversal can clearly be seen. }
\label{fig:RecoilVmin} 
\end{figure}

The $10\%$ deviations in the recoil spectra are larger than the
typical deviations expected due to the annual modulation. Therefore
these deviations from the Gaussian model can also influence the
annual modulation signal. In Fig.~\ref{fig:ModulationAll} (top row) 
we plot the dimensionless recoil rate $(R(t)- \langle R \rangle ) / \langle R \rangle$ 
of all local $2$~kpc boxes at $\sim 8$~kpc from the centre of Aq-A-1 (left),where $\langle R \rangle$ is 
the annual mean recoil. 
We have assumed $v_{\rm min}=300$~km/s for all plots in this figure.  
The maximum can clearly be seen around the month of June. The plot on 
the right in Fig.~\ref{fig:ModulationAll} shows the modulation parameter 
space that we define by the day of maximum amplitude ($x$-axis) and the 
maximum modulation amplitude of the recoil rate over the year defined as $(R_{\rm max}-\langle R \rangle
)/\langle R \rangle$ ($y$-axis). The bottom row of Fig.~\ref{fig:ModulationAll}
shows the maximum amplitude (left) and day of maximum (right) for all
level-2 halos (solid red) and the corresponding best-fit
multivariate Gaussian model (dashed black).

Comparing the Gaussian median values to the box median values one 
can see that the day of maximum amplitude does not deviate 
significantly from that predicted for a multivariate Gaussian; in particular there is 
no clear trend visible over the halo sample. On the other hand, the 
median amplitude in all halos is slightly higher than expected based on the 
Gaussian sample for $v_{\rm min}=300$~km/s. 

The amplitude differences for various $v_{\rm min}$ values are shown
in Fig.~\ref{fig:RecoilVmin}.  Here we calculated the maximum
amplitude and day of maximum for different $v_{\rm min}$ values for
all level-2 halos. The amplitude plot (top) shows the difference 
between the maximum relative modulation amplitude observed in the simulation
and that expected for the best-fit multivariate Gaussian model.
The maximum amplitude $v_{\rm min}$-dependence is similar for the 
six halos. Since only the velocity distribution enters into the 
recoil calculation, this similarity is due to the fact that 
deviations of the halo velocity distribution from the Gaussian model 
are also quite similar for all six halos. The bottom plot of
Fig.~\ref{fig:RecoilVmin} shows the day of maximum amplitude is
well predicted by the multivariate Gaussian for all 
halos.  The sharp transition in the day of maximum is due to the
well-known phase-reversal effect \citep{1988ARNPS..38..751P}.
We checked that the subhalo dominated box in Aq-B-2, where by chance 
about $60\%$ of the box mass is in a single subhalo, leads to a very 
different modulation signal. The day of maximum in that case shifts 
about $100$ days from the Gaussian distribution. We note
that although the subhalo mass fraction in this particular
box is high, the subhalo volume fraction is tiny, so even within
this box, almost all observers would see the smooth regular signal. 

\begin{figure*}
\center{
\includegraphics[width=0.4\textwidth]{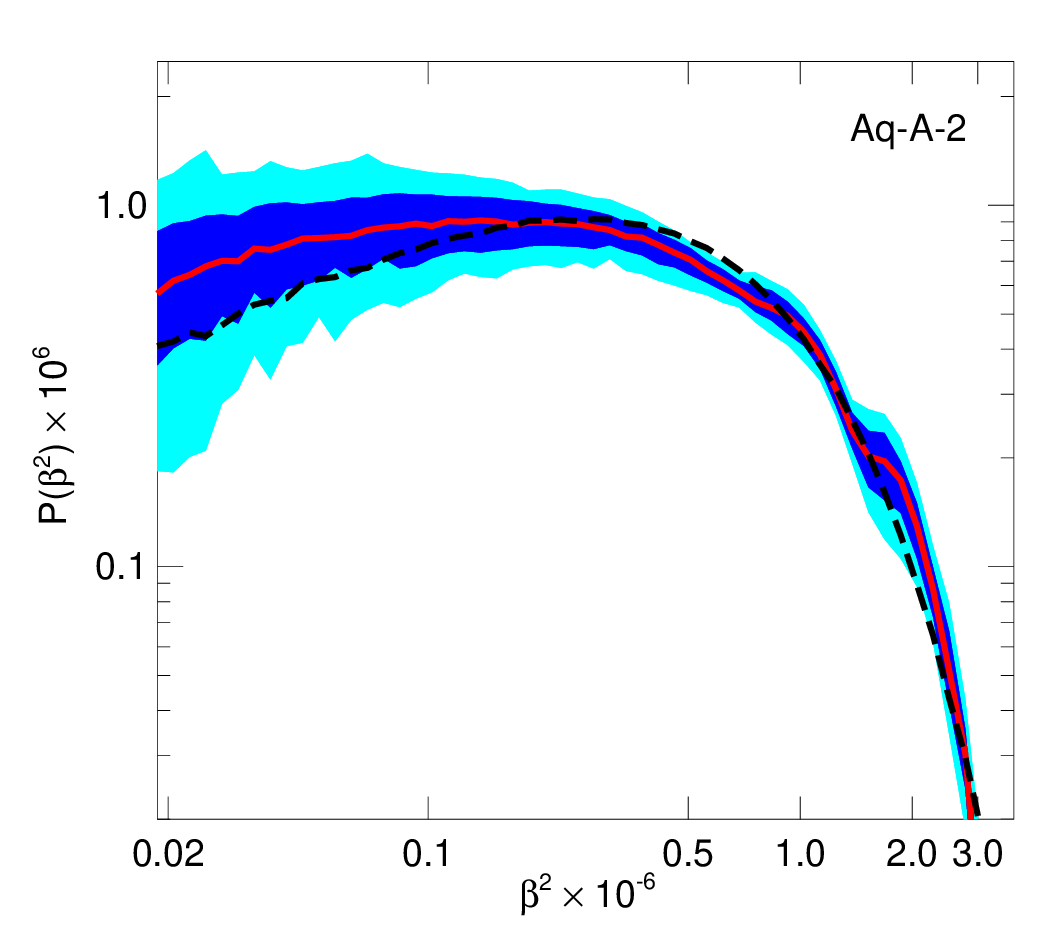}
\includegraphics[width=0.4\textwidth]{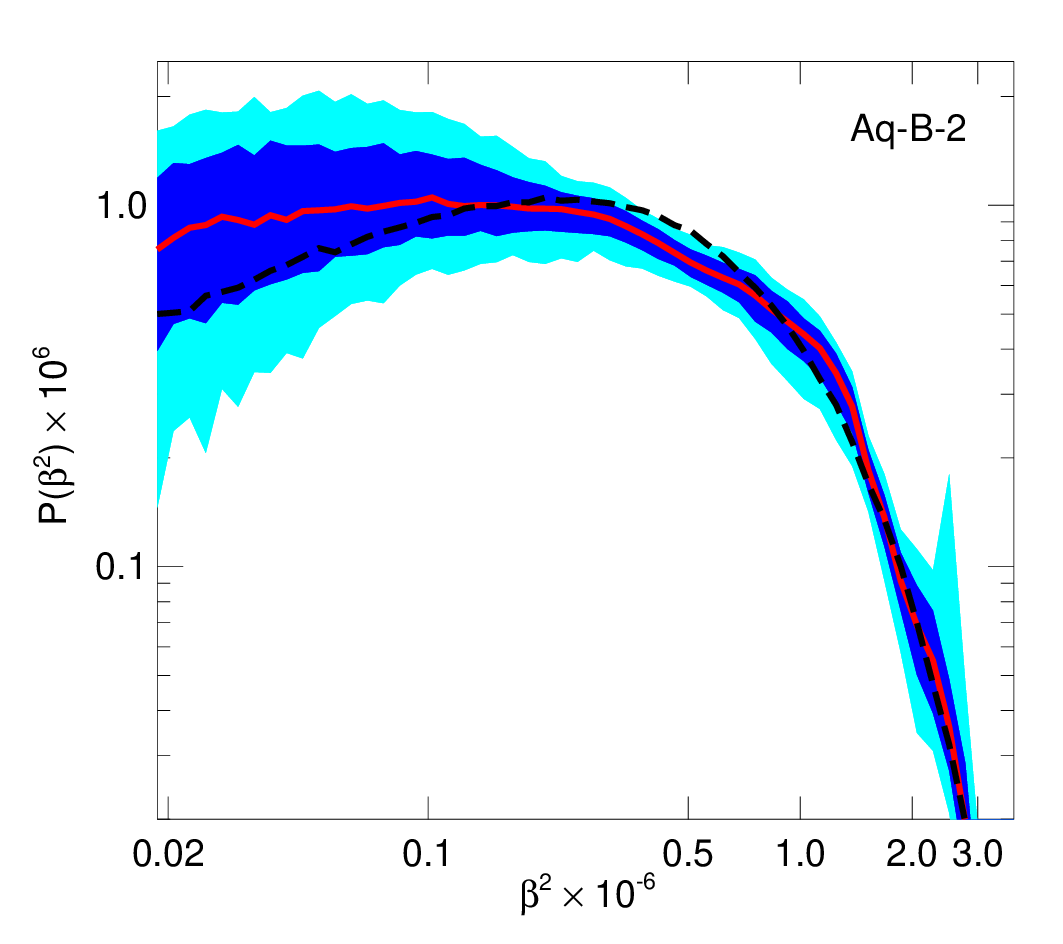}
\includegraphics[width=0.4\textwidth]{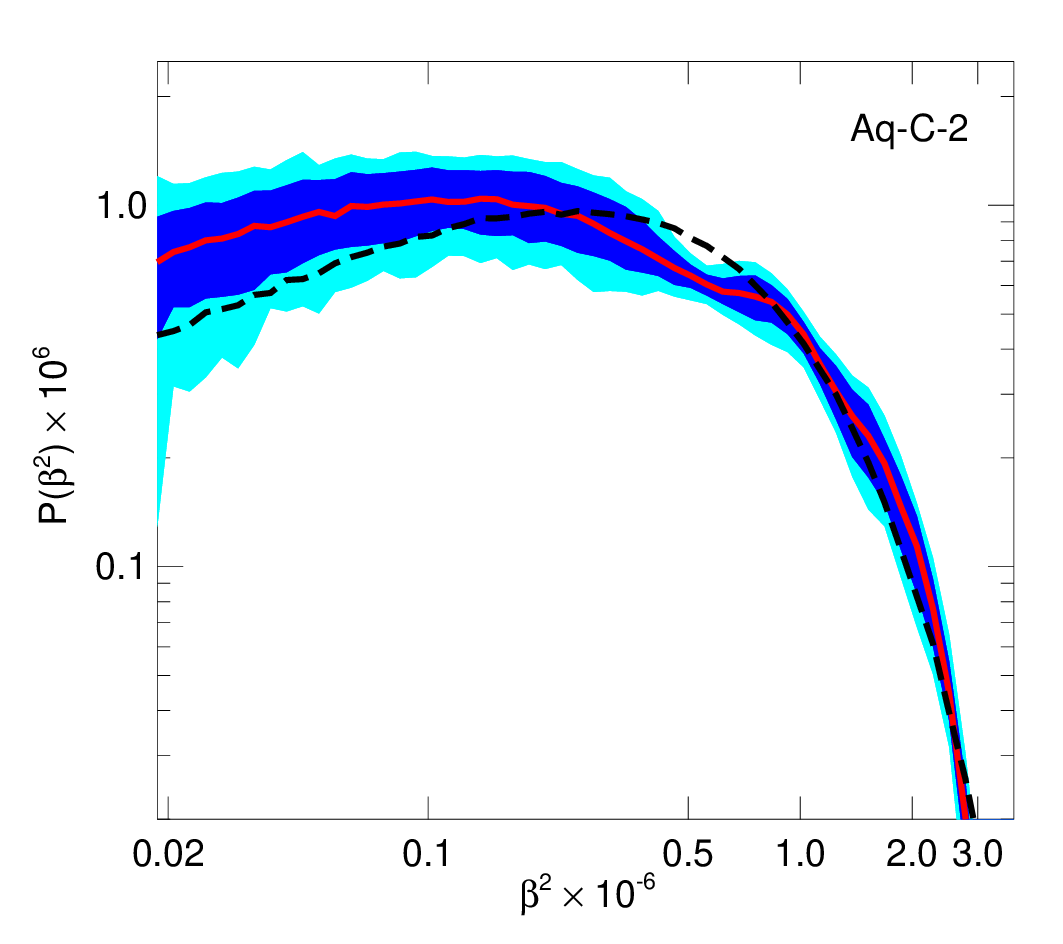}
\includegraphics[width=0.4\textwidth]{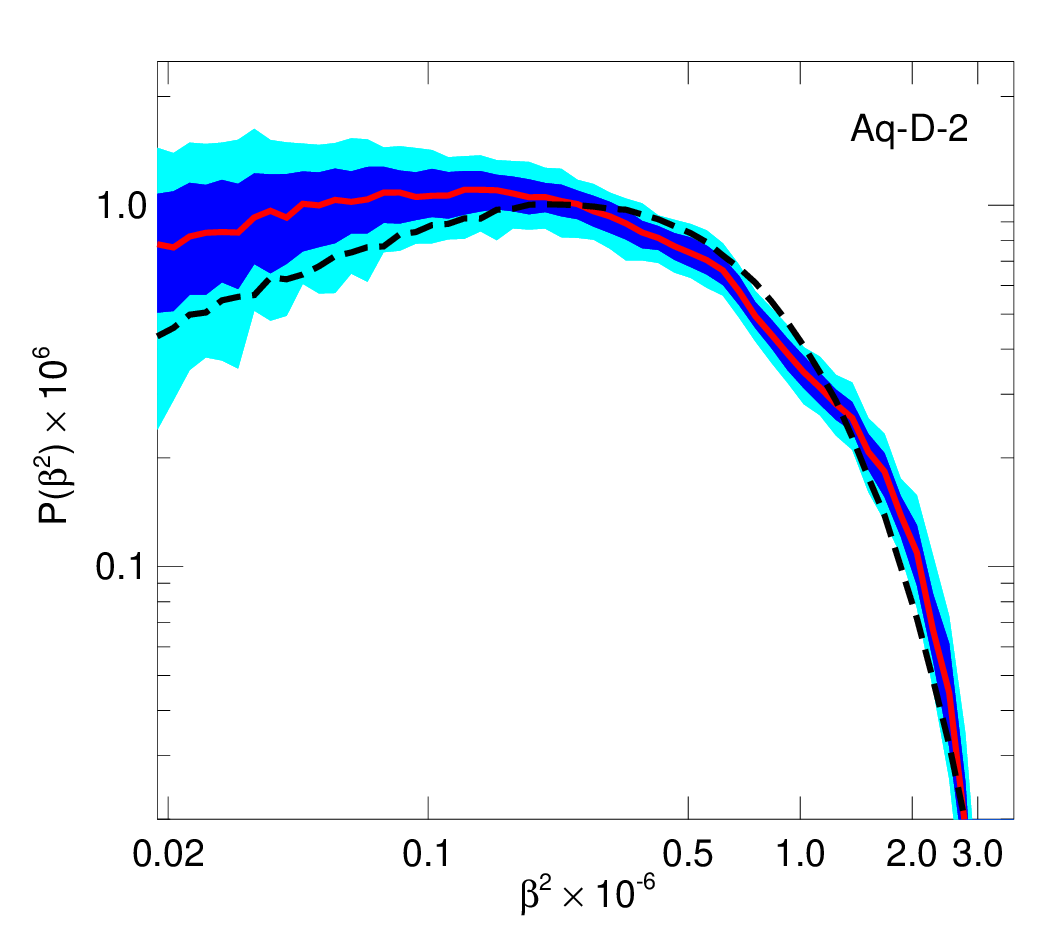}
\includegraphics[width=0.4\textwidth]{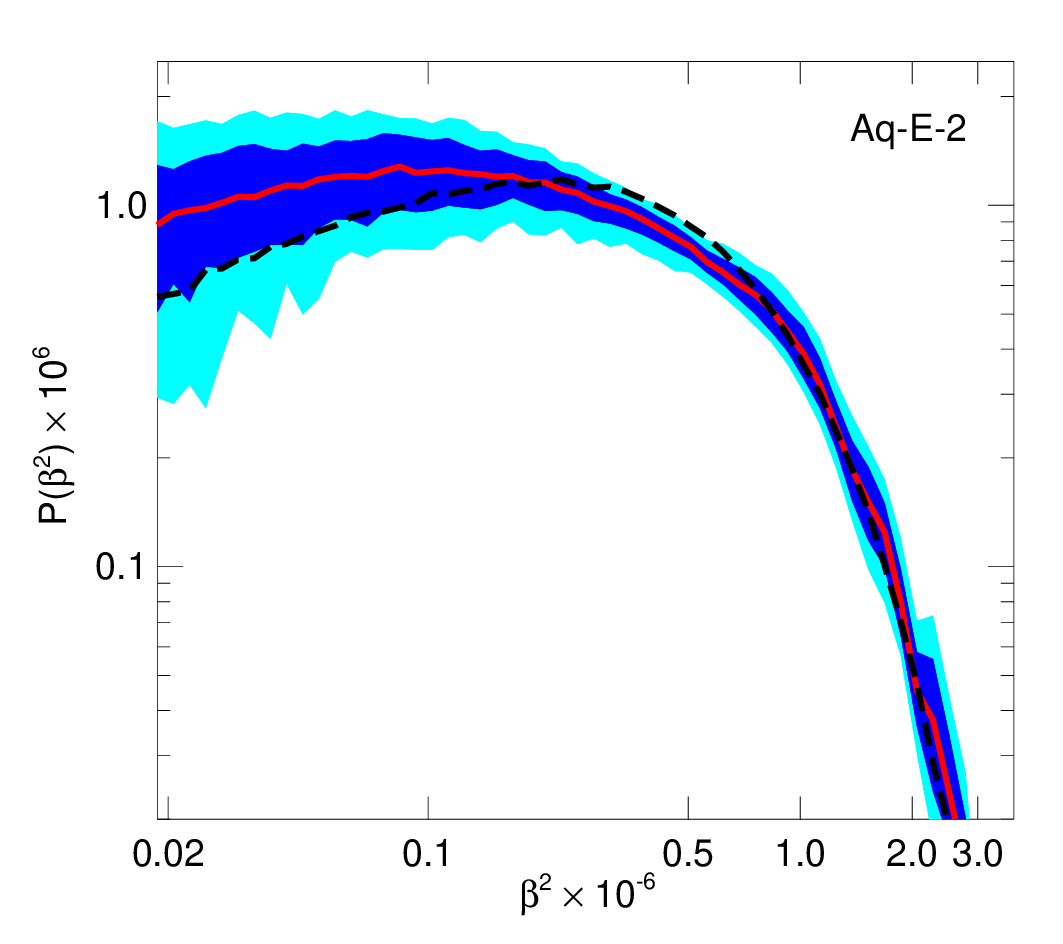}
\includegraphics[width=0.4\textwidth]{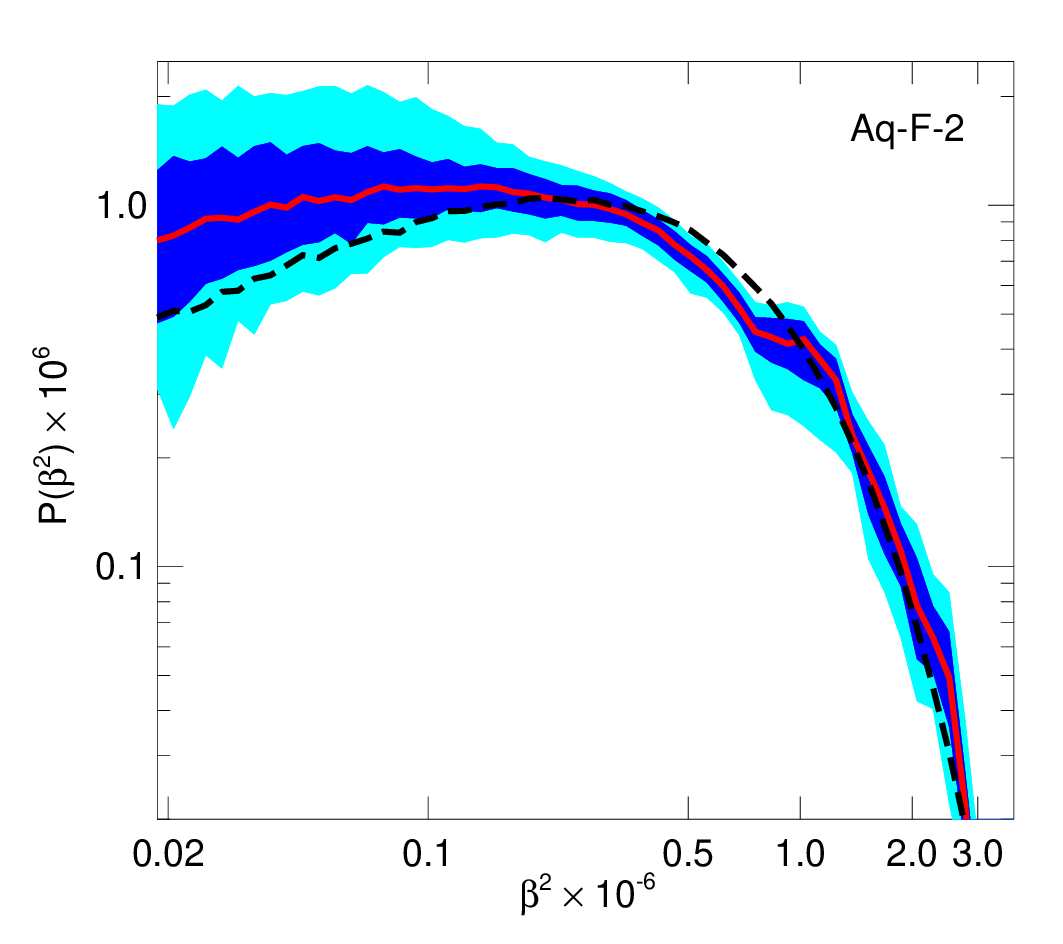}
}
\caption{Axion spectra of level-2 halos for all $2$~kpc boxes with halocentric
 distance between $7$ and $9$~kpc. Rescaling the
 $x$-axis according to $x \rightarrow 241.8~m_a~(1 + 1/2 x)$ for an
 axion mass $m_a$ in $\mu$eV yields the $x$-axis in MHz. The $y$-axis 
 is proportional to the power $P$ developed in the detector cavity.  
 Therefore the panels show the predicted frequency spectra expected 
 for an axion search experiment like ADMX. These spectra can be 
 reasonably described by a multivariate Gaussian but significant differences 
 remain. The maximum in the power is at lower frequencies in the
 simulation than in the Gaussian model. The
 bumps already found in the velocity and energy distribution
 are clearly visible in these spectra. In all halos the power at low
 frequencies is higher than expected from the Gaussian model.
 The line and contour scheme are the same as in Fig.~\ref{fig:BoxVelA}.}
\label{fig:StackedAxionB} 
\end{figure*}

Although most of the direct experiments currently search for WIMPs the
axion provides another promising candidate for CDM. It arises from the
Peccei-Quinn solution to the strong CP-problem. One axion detection
scheme is based on using the axion-electromagnetic coupling to induce
resonant conversions of axions to photons in the microwave frequency
range.  Galactic axions have non-relativistic velocities
($\beta=v/c \sim 10^{-3}$) and the axion-to-photon conversion process
conserves energy, so that the frequency of converted photons can be
written as:
\begin{equation}
\nu_a = \nu_a^0 + \Delta \nu_a = 
241.8 \left(\frac{m_a}{1 \mu {\rm eV}/c^2}\right)\left(1+\frac{1}{2} \beta^2\right) \rm{MHz} 
\end{equation}
where $m_a$ is the axion mass that lies between $10^{-6}$~eV/$c^2$ and
$10^{-3}$~eV/$c^2$. $5\mu$eV axions would therefore convert
into $\nu_a^0 \cong 1200$~MHz photons with an upward spread of
$\Delta \sim  \cong 2$~kHz  due to their kinetic
energy. An advantage of axion detection compared to WIMP searches is
the fact that it is directly sensitive to the energy rather
than to the integral over the velocity distribution.  The power $P$,
developed in the axion search cavity due to resonant axion-photon
conversion can be written as \citep{1983PhRvL..51.1415S}:
\begin{equation}
P=\mathcal{P} ~ \rho_a(\nu_{\rm cavity})
\end{equation}
where $\mathcal{P}$ encapsulates the experimental properties (cavity
volume, magnetic field, quality factor) and particle physics
properties (model dependent coupling parameter, axion mass). The only
astrophysical input is the local density $\rho_a(\nu_{\rm cavity})$ of
axions with energies corresponding to the cavity frequency. For simplicity
we set $\mathcal{P}=1$. We can produce axion spectra from our simulations 
by taking a local volume element (a box) and computing the distribution 
of kinetic energies $K$ of the particles found in this location. The number 
of particles with a given $K$ is then directly proportional to $\rho_a$ at this
frequency, and so to the power in the frequency bin.

To make the results independent of axion mass and other experimental
properties we present histograms of $\beta^2$ normalised to one. For a
given axion mass $m_a$ (in $\mu {\rm eV}/c^2$) the $x$-axis must be
transformed according to $x \rightarrow 241.8~m_a~(1 + 1/2 x) $ to get
the corresponding frequencies in MHz.  

A long-running axion search experiment is ADMX at LLNL
\citep{1996NuPhS..51..209H}.  It has channels at medium (MR) 
and high resolution (HR). The latter has a
frequency resolution of about $0.02$~Hz.  For $\nu_a^0=500$~MHz and an
axion velocity of $v=200$~km/s this translates into a velocity width
of only $0.018$~km/s\footnote{For non-relativistic motion we can write
$\rm{d}v=(c^2/v)~(\rm{d}\nu/\nu_a^0)$.}. Our numerical resolution
prevents us from predicting the behaviour on such small
scales. For wider bin searches and especially for the medium
resolution (MR) channel ($125$~Hz corresponding to a typical velocity
spread of about $100$~km/s) we can, however make reliable predictions by 
binning particles with respect to $\beta^2$.

In Fig.~\ref{fig:StackedAxionB} we show axion spectra for all
level-2 halos\footnote{We neglect the effects of the Earth's motion
when constructing the spectra since here our focus is on the general
spectral shape. This motion typically leads to a shift of about
$100$~Hz due to annual modulation and a daily shift of about 
$1$~Hz due to the Earth's rotation \citep{2005PhRvL..95i1304D}.}. 
In a broad sense, the spectra obtained from our simulations
look similar to those of multivariate Gaussian
models. However there are a number of differences. For example, the
peak power is shifted to lower frequencies. The Gaussian distribution
is also a poor description of the spectrum at low
frequencies.  In all halos the power at low and high frequencies is
higher then expected from a multivariate Gaussian model. This
effect is quite small for high frequencies but very significant for
low frequencies. The higher power at low frequencies can be understood 
from the velocity distributions in Fig.~\ref{fig:BoxVelAll}. 
In Aq-B-2 the subhalo dominated box that was seen in Fig.~\ref{fig:BoxVelAll} 
is clearly visible as a peak in the power spectrum at high frequency. The bumps in the velocity 
distribution also result in quite significant features in the axion spectra 
that might be visible in the MR channel given enough signal-to-noise.

\section{Conclusion}

We have characterised the local phase-space distribution of dark matter using
the recently published ultra-high resolution simulations of the
Aquarius Project. Our study provides new insights relevant
to searches for the elusive CDM particles.
This results from the unprecedented resolution and convergence (in a dynamical
sense) of our simulations, as well as from the fact that they provide a
sample of six Milky Way-like dark matter halos.

We have measured the probability distribution function of the DM mass
density between $6$ and $12$~kpc from the centre of the halo, finding
it to be made up of two components: a truly smooth distribution
which scatters around the mean on ellipsoidal shells by less than
$5\%$ in all the halos of our sample, and a high-density tail associated
with subhalos. The smooth DM component dominates
the local DM distribution.  With $99.9\%$ confidence we can say that
the Sun lies in a region where the density departs from the mean on
ellipsoidal shells by less then $15\%$.  Experimentalists can safely
adopt smooth models to estimate the DM density near the Sun.

We find that the local velocity distribution is also expected to be very
smooth, with no sign of massive streams or subhalo contributions.  The
standard assumption of a Maxwellian velocity distribution is not
correct for our halos, because the velocity distribution is clearly anisotropic.
The velocity ellipsoid at each point aligns very well with the shape of the
halo. A better fit to the simulations is given by a multivariate
Gaussian. Even this description does not reproduce the exact shape of
the distributions perfectly. In particular the modulus of the velocity
vector shows marked deviations from such model predictions. 
Velocity distributions in our six different halos share common
features with respect to the multivariate Gaussian model: the 
low-velocity region is more populated in the simulation; the peak of the
simulation distribution is depressed compared to the Gaussian; at high
velocities there is typically an excess in the simulation distribution
compared to the best-fit multivariate Gaussian.  Furthermore the
velocity distribution shows features which are stable in time, are
reproduced from place to place within a given halo, but differ between
different halos. These are related to the formation history of each
individual halo.

The imprints in the modulus of the velocity vector reflect
features in the energy distribution. We explicitly show that the
phase-space distribution function as a function of energy 
contains characteristic wiggles. The amplitude of these wiggles with
respect to the average distribution function of our sample of six
halos rises from high to low-binding energies. After appropriate
scaling, the most bound part of the distribution function looks very
similar in all halos, suggesting a (nearly) universal shape.  The weakly
bound part of the distribution, on the other hand, can
deviate in any given halo by an order of magnitude from the mean.

We have used our simulations to predict detector signals for WIMP and
axion searches. We find that WIMP recoil spectra can deviate about
$10\%$ from the recoil rate expected from the best-fit multivariate
Gaussian model. The energy dependence of these deviations looks
similar in all six halos; especially at higher binding energies.  We
find that the annual modulation signal peaks around the same day as
expected from a multivariate Gaussian model with no clear trend over
our halo sample for varying recoil velocity thresholds.  The maximum
recoil modulation amplitude, on the other hand, shows a clear
threshold-dependent difference between the signal expected for a
multivariate Gaussian model and that estimated from the simulation.
We have also explored the expected signal for direct detection of
axions. We find the axion spectra to be smooth without any sign of
massive streams. The spectra show characteristic deviations from those
predicted by a multivariate Gaussian model; the power at low and high
frequencies is higher than expected. The most pronounced effect is
that the spectra peak at lower frequencies than predicted.  Since the
frequencies in the axion detector are directly proportional to the
kinetic energy of the axion particles, the bumps in the DM velocity
and energy distributions are also clearly visible in the axion spectra. All
the effects on the various detector signals are driven by differences
in the velocity distribution. Individual subhalos or streams do not
influence the detector signals however, since they are sub-dominant by
a large factor in all six halos.

Our study shows that, once direct dark matter detection has become
routine, the characterisation of the DM energy distribution
will provide unique insights into the assembly history of the Milky
Way halo. In the next decade, a new field may emerge, that of
``dark matter astronomy''.

\section*{Acknowledgements}
The simulations for the Aquarius Project were carried out at the
Leibniz Computing Center, Garching, Germany, at the Computing Centre
of the Max-Planck-Society in Garching, at the Institute for
Computational Cosmology in Durham, and on the `STELLA' supercomputer
of the LOFAR experiment at the University of Groningen. The analysis
for this work was mainly done on the OPA and VIP computing clusters at
the Computing Centre of the Max-Planck-Society in Garching.  MV
acknowledges the Kapteyn Astronomical Institute in Groningen for a
productive atmosphere where most of this work was done. MV thanks Anne
Green for helpful discussions. AH acknowledges financial support from
NOVA and NWO. This research was supported by the DFG cluster of excellence ``Origin and
Structure of the Universe''.

\label{lastpage}
\end{document}